\documentclass[journal]{vgtc}                     

\onlineid{1576}

\vgtccategory{Research}

\vgtcpapertype{evaluation}

\title{Haptic Stylus vs. Handheld Controllers: A Comparative Study for Surface Visualization Interactions}

\author{%
  \authororcid{Hamza Afzaal}{0009-0004-8066-0523},
  and \authororcid{Usman Alim}{0000-0003-4834-2475}
}

\authorfooter{
  \item
  	Hamza Afzaal and Usman R. Alim are with University of Calgary. E-mails: \{hamza.afzaal, ualim\}@ucalgary.ca
}

\newcommand{\hamza}[1]{{\color{brown} Hamza: [{#1}]}}

\abstract{%
Surface visualizations are essential in analyzing three-dimensional spatiotemporal phenomena. Given its ability to provide enhanced spatial perception and scene maneuverability, virtual reality (VR) is an essential medium for surface visualization and interaction tasks. Such tasks primarily rely on visual cues that require an unoccluded view of the surface region under consideration. Haptic force feedback is a tangible interaction modality that alleviates the reliance on visual-only cues by allowing a direct physical sensation of the surface. In this paper, we evaluate the use of a force-based haptic stylus compared to handheld VR controllers via a between-subjects user study involving fundamental interaction tasks performed on surface visualizations. Keeping a consistent visual design across both modalities, our study incorporates tasks that require the localization of the highest, lowest, and random points on surfaces; and tasks that focus on brushing curves on surfaces with varying complexity and occlusion levels. Our findings show that participants took longer to brush curves using the haptic modality but could draw smoother curves compared to the handheld controllers. In contrast, haptics was faster in point localization, but the accuracy depended on the visual cues and occlusions associated with the tasks. Finally, we discuss participant feedback on using haptic force feedback as a tangible input modality and share takeaways to help outline design strategies for using haptics-based tangible inputs for surface visualization and interaction tasks. 
}


\teaser{
  \centering
  \includegraphics[width=\linewidth, trim={0 6.2cm 1.6cm 0}, clip]{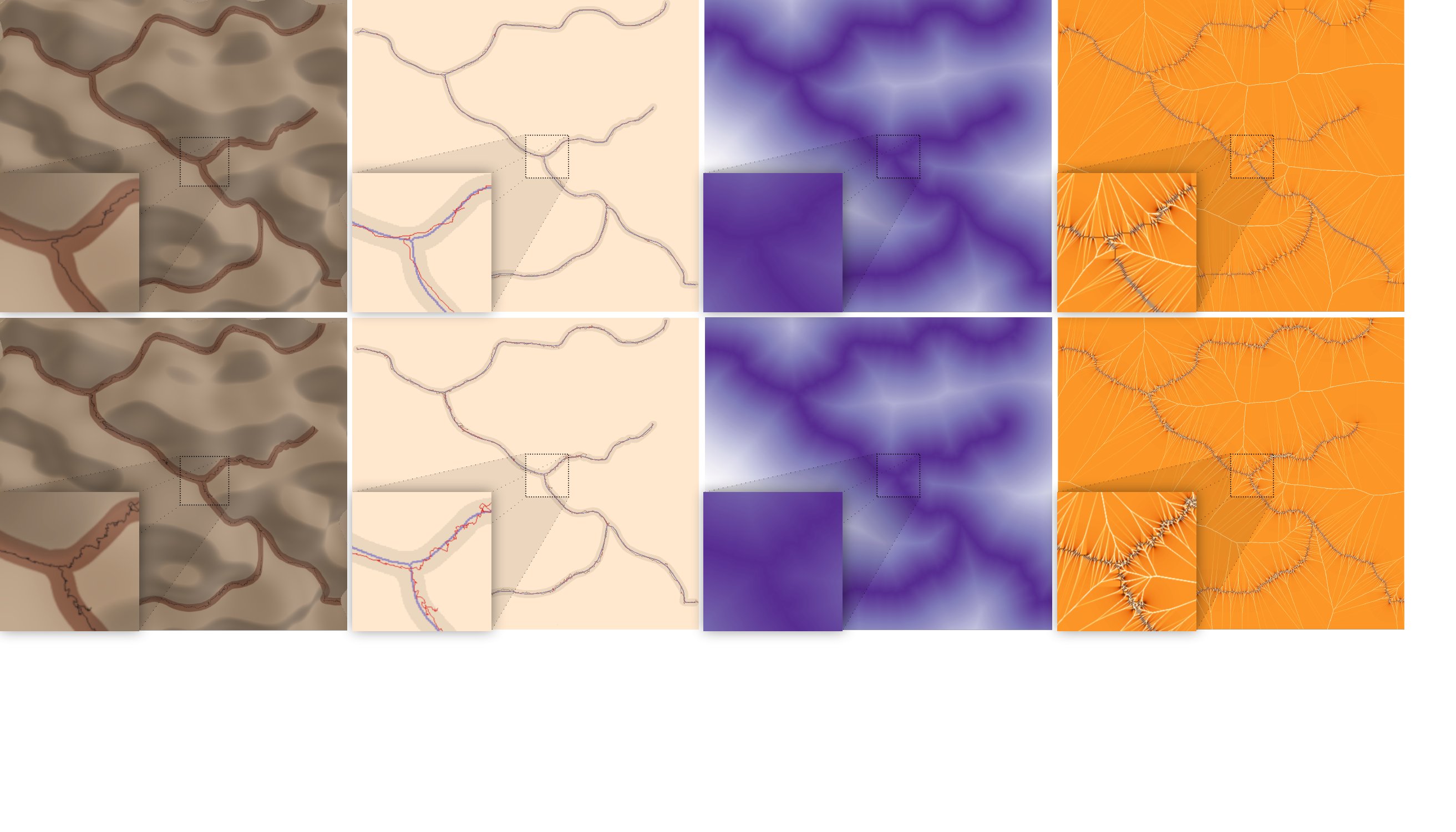}
  \caption{%
  Participants brushed curves on a surface in VR using either a haptic stylus (top) or a handheld controller (bottom). Left to right: visualization of the curves on top of the surface, texture image showing the bands participants were asked to follow and the brushed curves, Euclidean distance transform of the brushed curves in texture space, and the Laplacian of the Euclidean distance transform. Observe that particpants using the haptic stylus brushed smoother curves compared to participants using a handheld controller.%
  }
  \label{fig:teaser}
}

\graphicspath{{figs/}{figures/}{pictures/}{images/}{./}} 

\usepackage{tabu}                      
\usepackage{booktabs}                  
\usepackage{lipsum}                    
\usepackage{mwe}                       
\usepackage{multirow}                  
\usepackage{textcomp}
\usepackage{amsmath}

\begin{document}
\definecolor{point_color}{RGB}{153, 52, 4}
\definecolor{brush_color}{RGB}{84, 39, 143}
\definecolor{sig_color}{RGB}{150, 150, 150}
\newcommand{\pointc}[1]{{\textcolor{point_color}{\texttt{{#1}}}}}
\newcommand{\brushc}[1]{{\textcolor{brush_color}{\texttt{{#1}}}}}
\newcommand{\sigc}[1]{{\textcolor{sig_color}{\texttt{{#1}}}}}


\maketitle
\section{Introduction} 
\label{sec:intro}

Surface visualizations are pivotal for analyzing, exploring, and interpreting three-dimensional shapes and properties from medical imaging, computer-aided design, and geospatial datasets\cite{1_sbfv_Edmunds, 2_rtIsoSurface_westermann, 8_GeoSpatial_Mills, 21_silhoulettes_vol_tietjen, 22_illustrative_rendering_Lawonn}. Interpreting such visualizations requires an understanding of the surface's depth, distance, and orientation \cite{3_precep_vis_surf_gibson, 4_view_dir_tex_orient_sweet}. Virtual reality (VR) has improved the comprehension of these visualizations \cite{7_survey_imm_analytics_fonnet} by providing visual cues such as field of regard, stereoscopy, and head tracking for better viewing angles and depth perception \cite{5_vr_fidelity_laha, 6_immersion_analytics_laha}. Still, perception and interaction with surface visualizations in virtual reality depend on some intrinsic properties like shading, lighting and texturing to better understand the surface's depth and shape \cite{4_view_dir_tex_orient_sweet, 10_spatial_perception_lawonn}. Moreover, surface visualizations inherit occlusion problems due to region overlaps, and hence the view has to be adjusted continuously to gain a better perspective of the entire visualization \cite{9_EvalVis_Meuschke, 12_occlusion_management_elmqvist, 20_survey_perceptual_vis_preim}. This continuous adjustment of the view affects the memorability aspect of the visualization, increasing the cognitive overhead needed to switch back-and-forth between regions of interest on a surface \cite{17_comparative_vis_ensemble_alabi}.

Researchers have explored different interaction modalities in combination with virtual reality to provide users with additional sensory stimuli for navigating and understanding surface visualizations \cite{13_state_of_the_art_besançon}. Haptification of 3D spatial data using a force-based input device is widely used in the scientific community to provide a touch sensation from the visualized data \cite{15_haptic_rendering_lin}. These tangible input devices provide an additional channel of information for perceiving visual cues like depth and shape of the visualizations by activating the lateral optical visuo-tactile area of the brain, hence complementing the interpretation of the visualizations \cite{14_seeing_through_touch_reiner}. Moreover, the physical perception of the surfaces with force-based haptics having six degrees of freedom (DoF) facilitates the interaction with occluded and cluttered regions \cite{13_state_of_the_art_besançon}.

Given the importance of surface visualizations in the scientific community and the problems outlined with the visual-only modes of interactions, our primary goal in this paper is to investigate how the improved depth and shape perception provided by the combination of VR and force-based haptics improves user performance on interactive surface visualization tasks. To the best of our knowledge, this is the first paper that presents a comparative study contrasting visual-only and visual-with-haptics modes for surface interaction tasks.   
To accomplish our goal, we designed and conducted a quantitative user study to investigate the use of force-based haptics for interactions with surface visualizations as compared to using VR handheld controllers that provide no haptic feedback.
Our study employs common visual interaction tasks such as point localization and brushing curves on surface visualizations~\cite{18_haptic_lasso_zohu, 19_viewsheds_li}. Our findings elucidates scenarios where the force-based haptics assists interaction (see \cref{fig:teaser}) with surface visualizations and scenarios where visual-only cues are sufficient to accomplish the tasks.
As a secondary goal, we also discuss a set of design guidelines for future researchers on utilizing force-based haptics for interacting with various forms of surface-based visualizations. 

Our contributions can be summarized as follows: 
\begin{itemize}
    \item A quantitative \textbf{user study} comparing users' performance on six interaction tasks involving surface visualizations. A total of 40 participants were asked to localize the highest, lowest, and random points, and brush curves on surfaces with varying complexity. We interpret and report the results from our study and provide design guidelines for building interactive visualization systems that involve force-based haptics. 
    \item A set of \textbf{takeaways} and lessons learned from the user study based on feedback received from the participants.
\end{itemize}

\section{Related Work}\label{sec:related_work}
In this section, we review previously techniques for surface visualizations and discuss their broad-spectrum use cases and applications. We also explore different techniques for interaction that rely on visual cues from the surface. Furthermore, we review the use of tangible and hybrid input modalities for interacting with surface visualizations to establish a baseline for our surface visualization design strategy. 

\subsection{Importance of Surface Visualization} \label{sec:imp_surf_vis}
Researchers have employed different techniques for visualizing 3D data to understand its intricate aspects. For volumetric data, the marching cubes algorithm is commonly used to extract isosurfaces by localizing scalar values and performing triangulation on voxels~\cite{48_marching_cubes_lorensen}. Sunderland \textit{et al}. \cite{47_contour_surfaces_sunderland} presented a technique for building surface-based visualizations from 2D contours extracted from slices in volumetric data. 
The visualization of parametric surfaces is pivotal in visualizing mathematical models and is an essential component of Computer Aided Design \cite{51_procedural_gen_Kinnear}. Applications of surface visualizations are also found in geospatial data such as digital elevation models (DEM) \cite{dem_dsm_zohu, dtm_moore}.

Surface visualizations are not limited to visualizing scalar data or mathematical models, but are also used for presenting additional information through encodings. Multivariate visualizations show additional information on surfaces by encoding data using colors and textures, which benefits fluid simulation applications \cite{16_touching_data_flow_vis_englund}. Rocha \textit{et al}. proposed decal-based techniques for visualizing multivariate data on arbitrary surfaces~\cite{RochaDecalMaps,44_decal_lenses_rocha}. Tietjen \textit{et al}.~\cite{21_silhouettes_tietjen} used silhouettes, volumes, and surfaces to depict components in a volumetric dataset with greater detail. Ropinski \textit{et al}.~\cite{37_depth_angiography_ropinski} proposed a depth-based color encoding technique that heightens depth perception in surfaces such as blood vessels. Alabi \textit{et al}.~\cite{17_comparative_vis_ensemble_alabi} proposed a method to superimpose surfaces to find similarities and differences between ensembles. Lawonn \textit{et al}.~\cite{22_illustrative_rendering_Lawonn} review works that use illustrative techniques to improve the perception of surface-based visualizations.

\subsection{Interaction Techniques for Surface Visualization}
Surface visualizations are highly view-dependent, so researchers have explored interaction techniques spanning the spectrum of output systems ranging from traditional 2D displays to immersive environments such as VR. Ohnishi \textit{et al}.~\cite{45_virtual_interaction_surfaces_ohnishi} decoupled interaction with 3D surfaces shown in a 2D environment by proposing virtual interaction surfaces for pointing, placement, and texture mapping on the surfaces. Wang \textit{et al}.~\cite{18_2d_3d_input_output_wang} explored the use of 2D and 3D input and output devices to test the efficacy of interaction with 3D visualizations and reported the benefits of using each input and output mode. Laha \textit{et al}.~\cite{5_vr_fidelity_laha} conducted a study to gauge the fidelity of the VR environment in analyzing isosurfaces by gauging pattern recognition, feature searching, and spatial judgment of features on the surfaces. Meuschke \textit{et al}.~\cite{9_EvalVis_Meuschke} conducted a perceptual study for identifying shapes and depths on surface visualizations. 
Song \textit{et al}.~\cite{46_tangible_handheld_song} explored the use of an external tangible input device such as an iPad to select, slice, and annotate volumetric datasets. Usher \textit{et al}.~\cite{30_neuron_tracing_usher} used VR handheld controllers for navigating and tracing neurons on a microscopy scan isosurface reconstruction of the visual cortex in primates. 

Another line of research has explored tangible inputs for performing complex manipulations on 3D data with simple real-world gestures and movements. In their book' Haptic Rendering,' Lin \textit{et al.}.~\cite{13_haptic_rendering_lin} talk about the algorithms and applications focusing on force-based haptics for interacting with surface visualizations. Paneels \textit{et al.}~\cite{10_haptic_data_vis_paneels} review the design strategies for incorporating haptics for interacting with surfaces. Avila \textit{et al.}~\cite{16_haptic_inter_methods_avila} used the PHANToM haptic interface for drawing, selecting, melting, and stamping color-based or physically induced features on surface visualizations. Basdogan \textit{et al.}~\cite{52_haptid_rendering_virtual_basdogan} discusses the applications of haptic interaction in immersive environments. Englund \textit{et al.}~\cite{16_touching_data_flow_vis_englund} used visuo-haptic visualizations to explore fluid flow by interacting and relaying sensory information through force-based haptics from isosurfaces. Corenthy \textit{et al.}~\cite{21_volume_haptic_consistent_iso_corenthy} explored using force-based haptics to interpret topology-consistent isosurfaces. Zhou \textit{et al.}~\cite{18_haptic_lasso_zohu} used 6-DoF from force-based haptic devices to draw lassos to select annotation regions on surfaces generated from diffusion tensor imaging. 

Haptics as a tangible mode of input has been widely used to perform interaction tasks with different forms of visualizations. As reviewed in a recent survey~\cite{13_state_of_the_art_besançon}, these haptic interactions range from vibrotactile, force feedback and ultrasonic feedback. Research has shown that perception of visualizations through physical touch improves cognition \cite{14_seeing_through_touch_reiner}. Researchers have also employed active assistance along with haptic force feedback to steer the user through annotation regions on 3D surfaces~\cite{spotlight_top}. Most work on the use of force-based haptics for surface visualizations has focused on techniques for generating haptic feedback for domain-specific data and tasks.  Our goal is to explore the domain-agnostic implications of using force-based haptics---\emph{without explicit steering assistance}---in comparison to a visual-only mode for commonly used interaction tasks in immersive surface visualizations.



\section{Design Choices} 
We now elaborate on the choices made for our surface visualizations and interactions, specifically differentiating between the choices for force-based haptics compared to the visual-only mode of interaction. 

\subsection{Visual Design}
The surface visualizations we employ involve various choices for the surface type, the visual design, and the rendering. These choices provide a common baseline that can be used to compare the two interaction modalities, and constitute 
the control variables of our study. We mainly focus on the generation of the type of surface models, the setup of the environment navigation controls, and the shading models. 

\begin{figure}[tb]
  \centering 
  \includegraphics[width=\columnwidth,trim={22cm 16cm 10cm 2cm},clip]{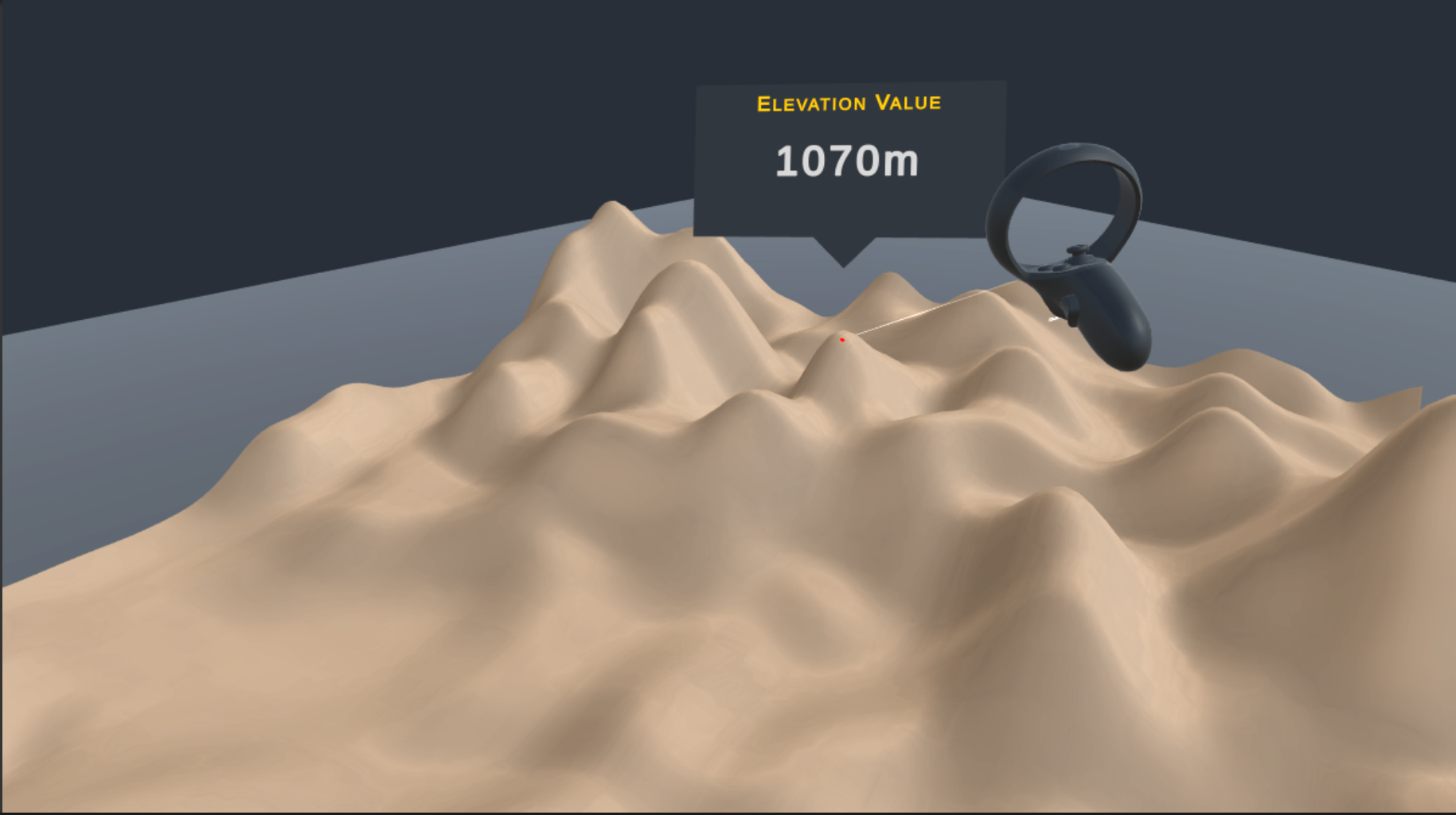}
  \caption{%
  	An example showing rendering on surface visualization and a VR handheld controller performing interaction with the visualization.
  }
  \label{fig:rendering}
\end{figure}
 
\subsubsection{Surface Models} \label{sec:surf_models}
Surfaces commonly used for visualization tasks can be broadly classified as either explicit or implicit. An explicit surface provides the surface points explicitly in 3D and typically employs a mesh representation. An implicit surface is implicitly given by a level-set of an underlying 3D scalar function. 
We used explicit surfaces as they are more prevalent in  interaction tasks that involve physical sensations~\cite{survey_haptic_rendering_laycock}.
They also allow us to easily map function values on top of the surfaces, which is important for multivariate data interactions on surface visualizations. 
We acquired surface models for our study in three ways: 
\begin{enumerate}
    \item\textit{Procedural surfaces} created by merging procedural textures such as Perlin noise \cite{perlin_noise}, 2D Gaussian and sinusoidal functions, as height maps to produce surfaces. We used randomized parameters and combined the textures using weighted blending to generate surfaces exhibiting a variety of features. 
    \item\textit{Hand sculpted surfaces} created using the sculpt tool for shaping a mesh in Blender. The sculpt tool provided precise control over the generation of specific features on the surface. We utilized sculpting to add surface features such as depressions.
    \item \emph{Object models} that were sourced from the surface annotation benchmark dataset of Chen \textit{et al.}~\cite{benchmark_3d_mesh_segment_chen}.
\end{enumerate}

In order to investigate interaction modalities in the presence of occlusion and clutter on surfaces, it is crucial to have a diversity of surface visualizations, as highlighted in \cref{sec:intro}. We drew inspiration for our surface models from existing literature (see \cref{sec:related_work}) on surface visualizations, which mainly utilize surfaces such as isosurfaces and parametric surfaces. Additionally, by generating surface models using procedural techniques, we have greater control over the variability of surface features, such as occlusion and clutter level, and can incorporate visual features like depth enhancement. We also employed both open and closed surface models as well as object models with varying topology to examine a broad spectrum of surface shapes. We then defined interaction tasks on the chosen surfaces as discussed in \cref{sec:int_des_rat}, and interacted with the surface visualizations using different modes of interaction.


\subsubsection{Navigation Controls}
When interacting with surface visualizations, it is crucial to effectively navigate the scene. While VR provides superior view manipulation capabilities through precise head tracking and field of view, appropriate controls are needed to facilitate the exploration of occluded and cluttered regions. We adopted an approach in line with existing literature \cite{worlds_in_wedges_nam}, where joysticks on handheld controllers were used to translate surface visualizations and rotate them about a central axis. The primary button was used for point selection, while the trigger button was used for brushing curves, and the joystick button was used for task submission. The navigation controls were configured on the non-dominant hand-held controller so that they were independent of the surface interaction controls that were mapped to the dominant hand. 


\subsubsection{Rendering} 
 Studies have shown the importance of proper shading techniques for visualizing surfaces as users rely heavily on depth and shape perception during interaction~\cite{4_view_dir_tex_orient_sweet}. In order to maintain consistency across visuals presented in our study, the surface  visualizations ertr rendered using a multi-pass strategy incorporating different shading techniques:
\begin{enumerate}
    \item \emph{Phong shading} to provide lighting cues for shape and depth.
    \item \emph{Screen-space ambient occlusion} (SSAO)\cite{ssao_mittring} to accentuate occluded regions and make the surface form more apparent. 
    \item \emph{Hill shading} \cite{hill_shading_kennelly}
    to aid the comprehension of the surface by highlighting the rapidly changing topography.
\end{enumerate} 
A sample rendering of a procedurally generated surface used in our study is shown in \cref{fig:rendering}. First, a base color was applied to the surface, followed by the shading techniques mentioned in the order above.
Furthermore, to assist 
users to quickly locate the point the interaction controller is aimed at, we rendered a laser pointer atop the visualizations to show the direction of the virtual proxies of the given interaction controller (see \cref{fig:rendering} and \cref{fig:inter_tasks}).


\subsection{Haptic Design}
The type of force-based haptic interaction we opted to use involves a set of design elements to allow a seamless interaction with the surface visualizations. We used a GeoMagic\textsuperscript{\sffamily\textregistered} Touch\textsuperscript{\sffamily\texttrademark} device to provide haptic force feedback with 6-DoF for assisting the interaction with surface visualization by providing an additional stimulus through touch. We used a passive mode of haptic force sensation which allows a free-hand movement on the surface for interactions. This design decision was made to evaluate the effectiveness of the haptic force feedback device without any additional active assistance. The haptic force from the device was limited to providing only a  physical sensation from the surface.  The haptic forces for physical touch are generated by the device in the direction of the surface normal when the virtual probe for the haptic device collides with the surface. We used the OpenHaptics\footnote{\href {https://www.3dsystems.com/haptics-devices/openhaptics}{https://www.3dsystems.com/haptics-devices (Accessed on 2023/03/30)}} library in Unity to configure the device as it provides basic surface interaction controls out of the box. For our study, the surface friction and marker speed were kept at the default settings. 

\subsection{Interaction Design} \label{sec:int_des_rat}

The proper selection of interactions for surface visualization is critical, and this research aims to evaluate the efficacy of force-based haptic interactions with surface visualizations. The goal is to synthesize interactions that highlight the limitations of visual-only modes of interaction. Besançon \textit{et al.}~\cite{13_state_of_the_art_besançon} have provided a taxonomy of interactions for 3D spatial data, emphasizing the importance of tangible inputs in 3D data selection and annotation tasks. These interactions include basic interactions with surface models that use points, curves, and regions to highlight features on surfaces. Due to the vast range of interaction types for 3D spatial data, we focus on two generic surface interactions: point localization and brushing curves.

\begin{figure*}[tbh]
  \centering
  \begin{subfigure}[b]{0.32\linewidth}
  	\centering
  	\includegraphics[width=\textwidth,trim={28cm 1.5cm 10cm 14cm},clip]{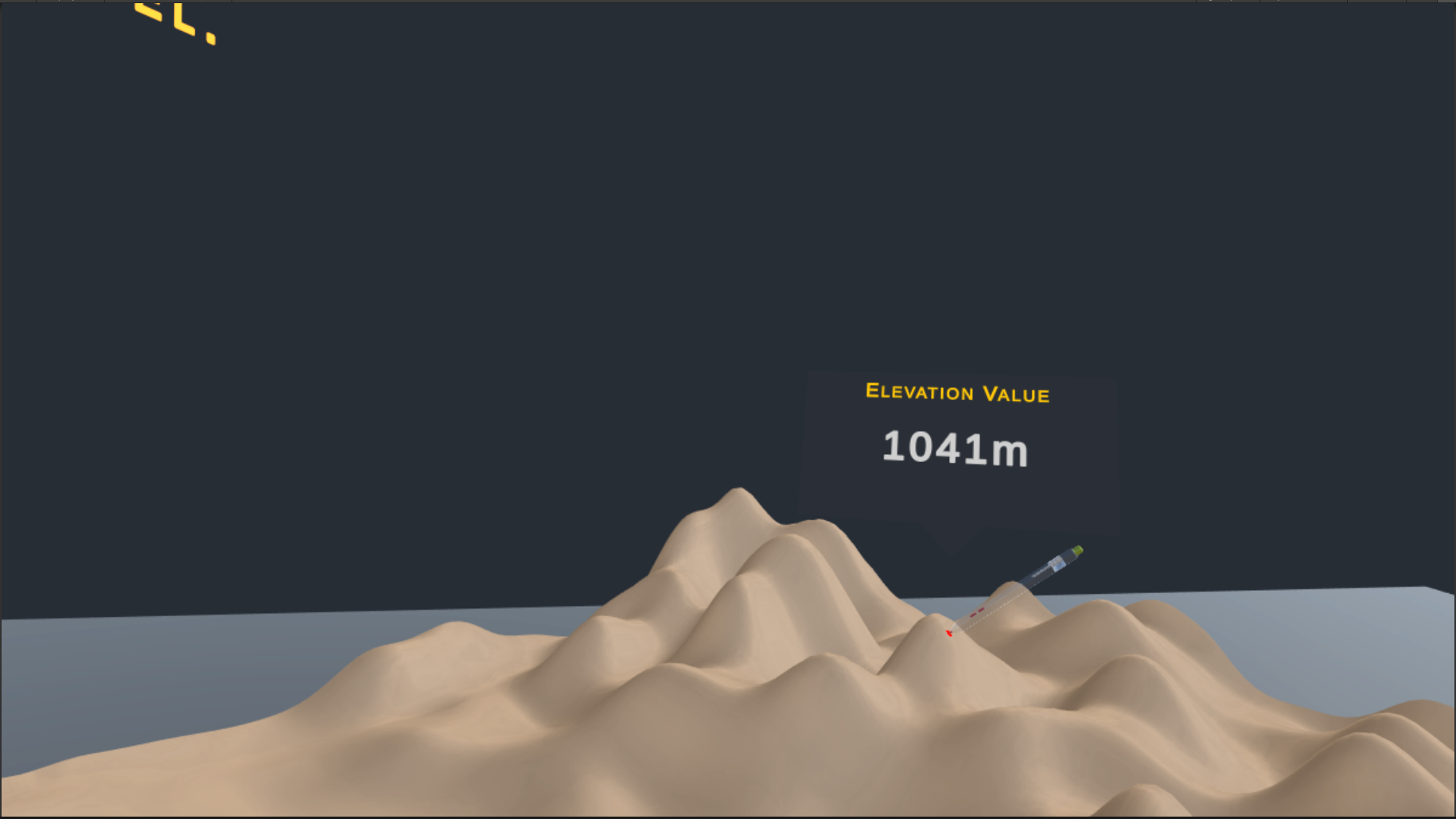}
  	\caption{Protrusion Point Localization}
  	\label{fig:protrusion}
  \end{subfigure}%
  \hfill%
  \begin{subfigure}[b]{0.32\linewidth}
  	\centering
  	\includegraphics[width=\textwidth,trim={24cm 1.5cm 14cm 14cm},clip]{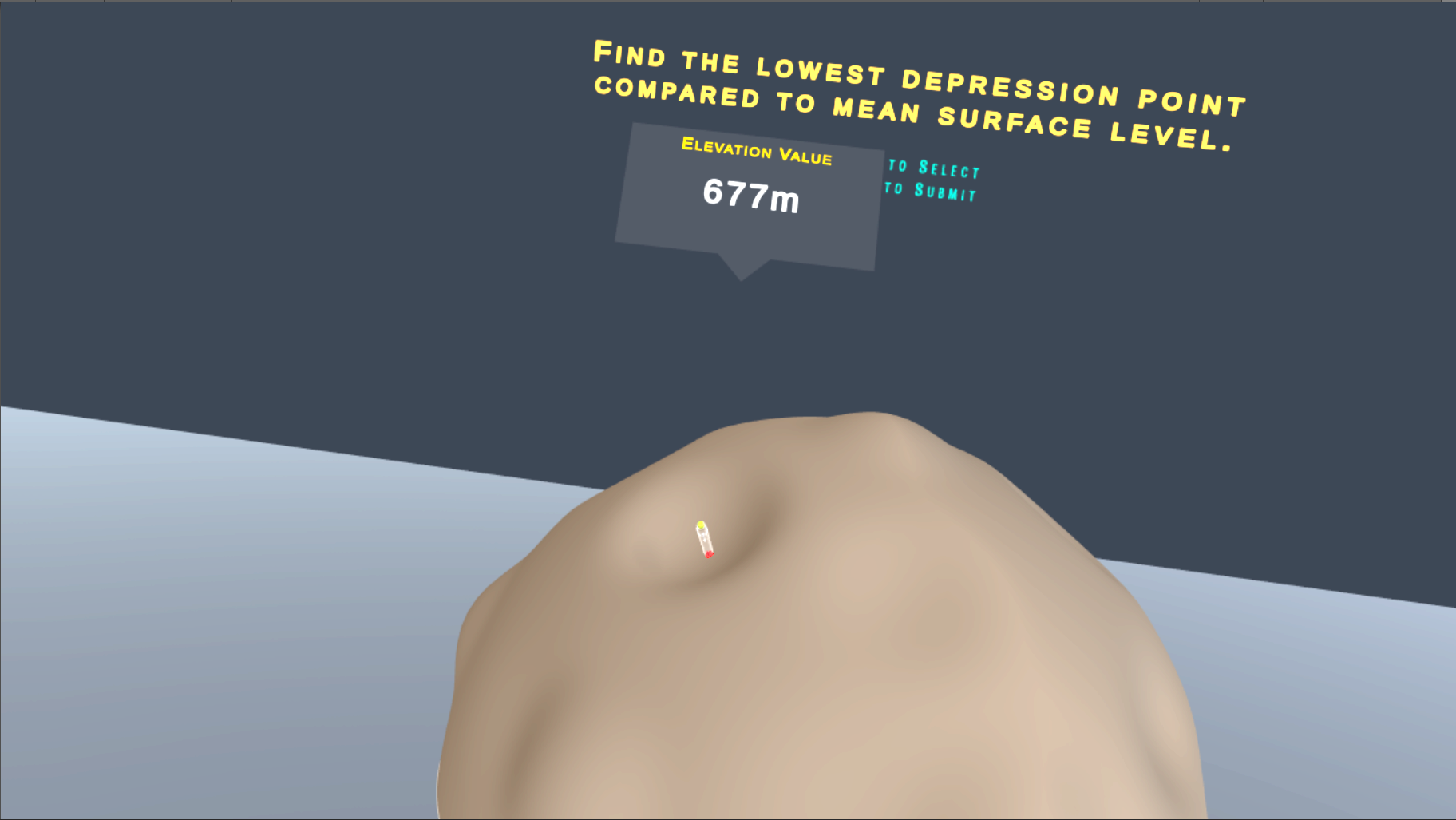}
  	\caption{Depression Point Localization}
  	\label{fig:depression}
  \end{subfigure}%
  \hfill
  \begin{subfigure}[b]{0.32\linewidth}
  	\centering
  	\includegraphics[width=\textwidth,trim={20cm 4.5cm 18cm 11cm},clip]{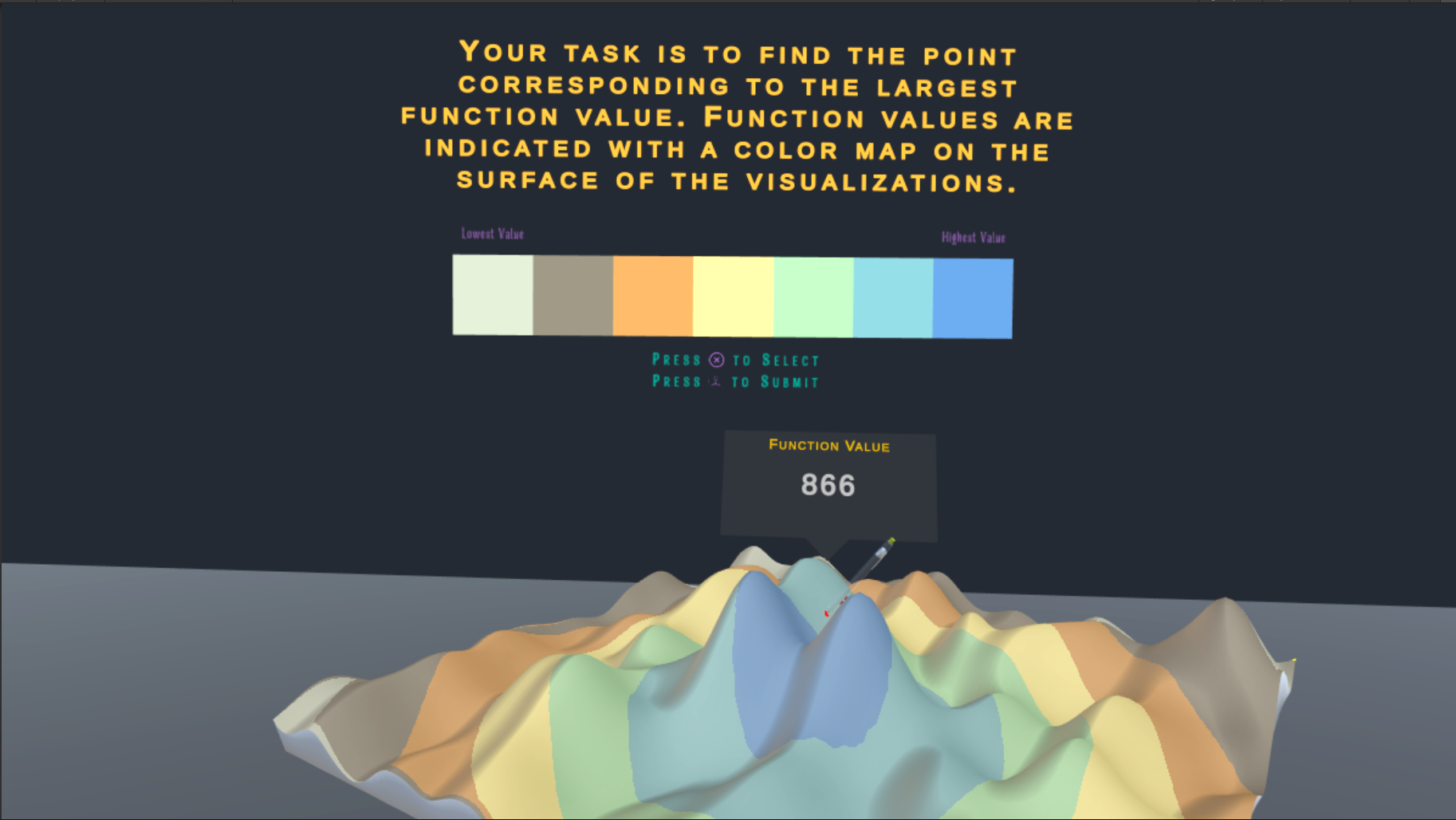}
  	\caption{Random Value Surface Point Localization}
  	\label{fig:randval}
  \end{subfigure}%
  \\
  \begin{subfigure}[b]{0.32\linewidth}
  	\centering
  	\includegraphics[width=\textwidth,trim={28cm 1cm 10cm 14cm},clip]{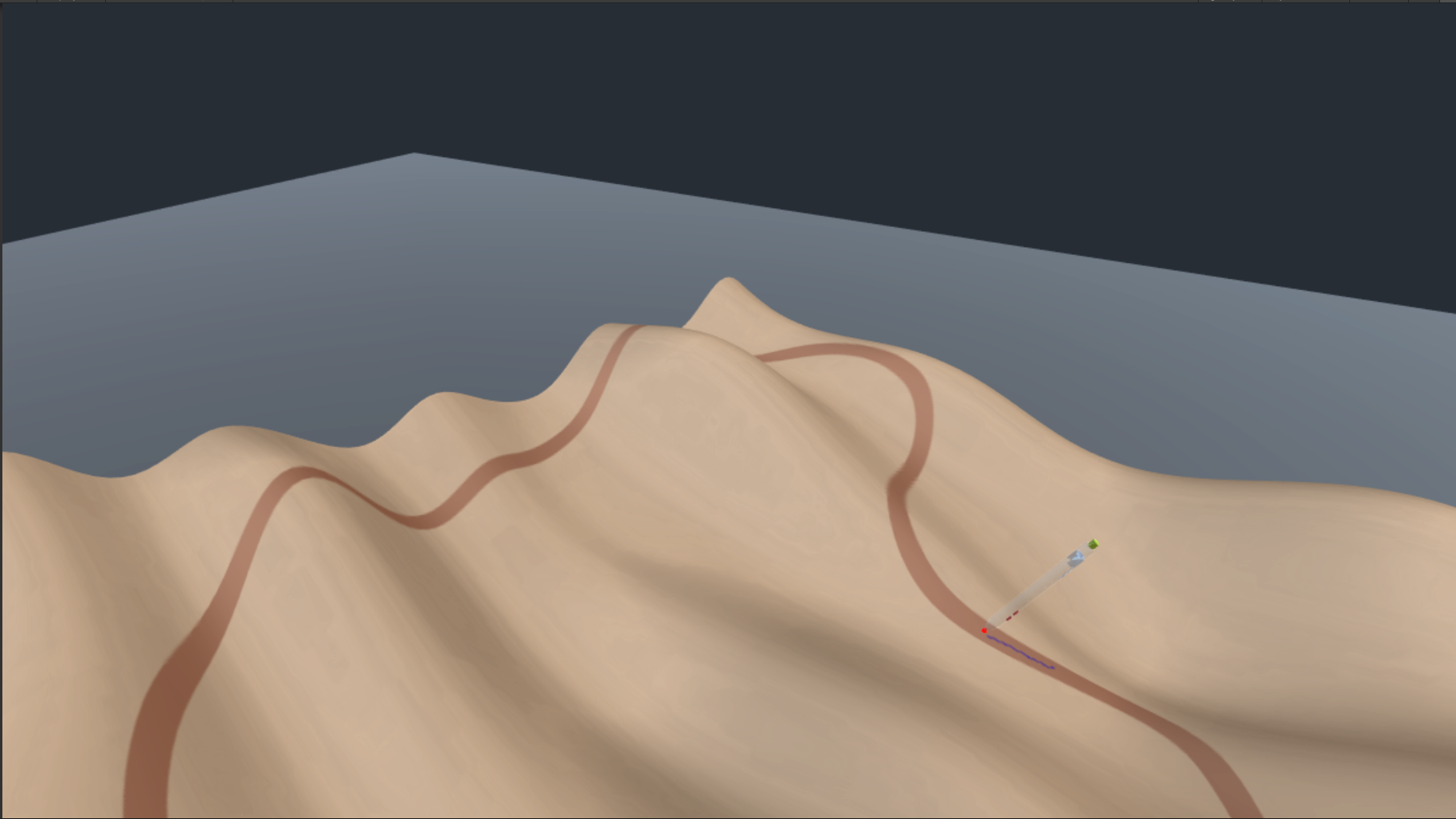}
  	\caption{Visual Contour Brushing}
  	\label{fig:contour}
  \end{subfigure}%
  \hfill%
  \begin{subfigure}[b]{0.32\linewidth}
  	\centering
  	\includegraphics[width=\textwidth,trim={28cm 1.5cm 10cm 14cm},clip]{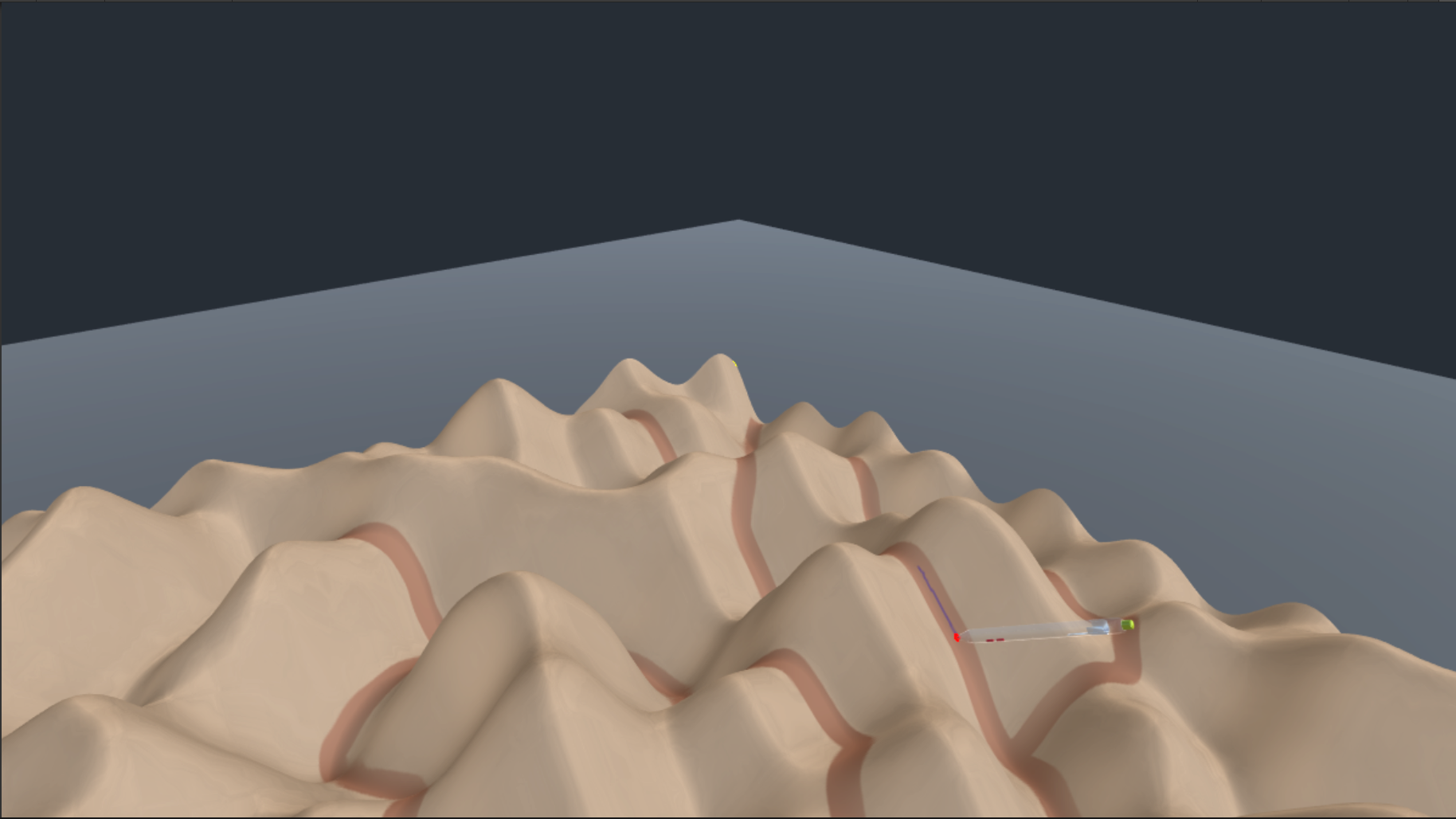}
  	\caption{Visually Marked Grooves Brushing}
  	\label{fig:groove}
  \end{subfigure}%
  \hfill
  \begin{subfigure}[b]{0.32\linewidth}
  	\centering
  	\includegraphics[width=\textwidth,trim={24cm 1.5cm 14cm 14cm},clip]{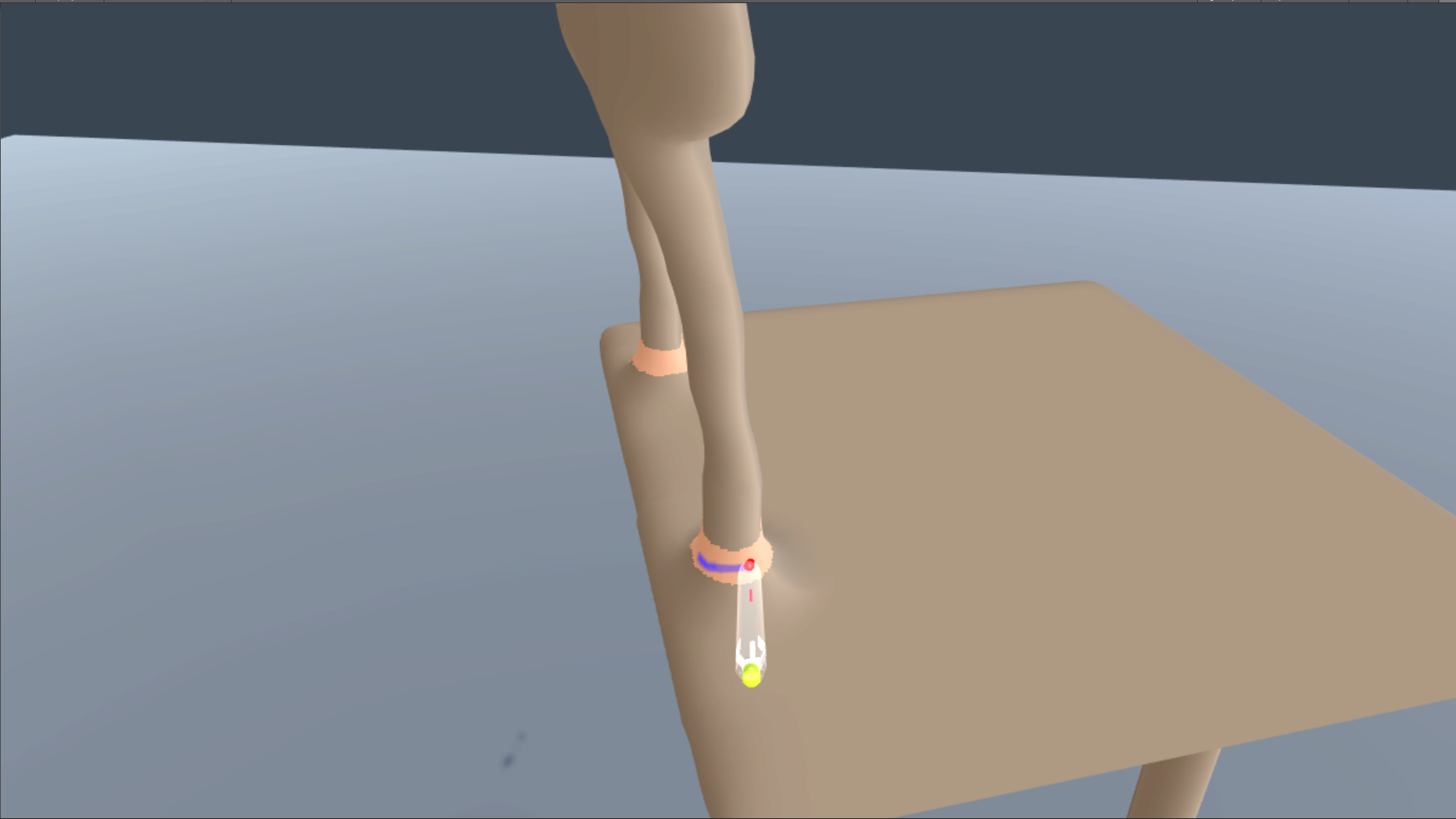}
  	\caption{Visually Annotated Regions Brushing}
  	\label{fig:annotation}
  \end{subfigure}%
  \subfigsCaption{All of the surface visualization interactions with the proxy for force-based haptic stylus.}
  \label{fig:inter_tasks}
\end{figure*}

\subsubsection{Point Localization} 
\emph{Point localization} entails the selection of a specific point on the surface to draw meaningful information either from inherent features of the surface or features that are indirectly mapped to the surface via functions, textures or decals~\cite{44_decal_lenses_rocha}, or a combination thereof.
Based on our outlined criteria, we identified three representative point localization interactions for surface visualizations. 

\textbf{\texttt{Protrusion Point Localization}} requires localizing the highest protruded point on a surface against its base surface level. For this, we generated procedural surfaces (see \cref{fig:protrusion}) with various protruded regions using method one described in~\cref{sec:surf_models}. The surface protrusion value at a point was computed by calculating its distance from the lowest point along the elevation axis
on the surface and scaling it to a numeric range of values that represents realistic elevation data. 

\textbf{\texttt{Depression Point Localization}} required localizing the lowest depression point on a surface against its base surface level. We sculpted a hemisphere-shaped surface model (see \Cref{fig:depression}) with various depression regions using method two described in \cref{sec:surf_models}. The surface depression value at a point was computed by calculating its distance from the center point of the hemispheric surface and scaling it to a realistic numeric range to assist interpretation.
A region that is depressed with respect to its surroundings provides different visuo-haptic cues as compared to a protruded region. We are therefore interested in investigating how these differences impact the localization of such points via a haptic stylus as compared to a VR handheld controller.

\textbf{\texttt{Random Value Point Localization}} required localizing the maximum function value as determined via a color map applied to the surface (see \Cref{fig:randval}).
Here, we generated synthetic functions by combining 2D anisotropic Gaussians with random means and variances. The resulting function was shown on the surface as a contour visualization using a diverging color map. 
It is necessary to provide visual assistance while interacting with these surface visualizations, as the function values mapped on the surface are independent of its shape and other properties. We generated surface models similar to the \texttt{Point Protrusion Localization} task. We used $7$ colors/contours with the highest range of values is mapped to blue. While the maximum point is located within the blue band, the user needs to search within the band to localize the maximum point. We are interested in observing differences between between a visual-only and a visuo-haptic interaction for the maximum point localization.

For all \texttt{Point Localization} tasks, we showed the corresponding surface values on a floating widget directly above the laser point that the interaction controller was aimed at~(see \cref{fig:rendering} and \cref{fig:inter_tasks}).


\subsubsection{Brushing Curves}
\emph{Brushing Curves} is another basic form of interaction performed on surface visualization. Common applications include feature or region demarcation and annotations for segmenting and clipping surfaces~\cite{13_state_of_the_art_besançon}. Curves are brushed on the surface by pointing directly towards the surface using various input modalities, such as a mouse, tablet, stylus, or a handheld controller, and then moving along the features on the surface. For our study, we identified three representative \texttt{Brushing Curves} interaction tasks for surface visualizations.

\textbf{\texttt{Visual Curve Brushing}} encompasses brushing a curve within a visually highlighted band on the surface. We generated procedural surfaces (see \Cref{fig:contour}) with various landmarks for adding occlusions and clutter using method one described in \cref{sec:surf_models}. Various curves were generated using 2D modelling tools; the curves were dilated to generate bands that were then mapped to 
to the shaded surface as a texture. The modelled curves therefore represent the medial axis of the bands. It is also worth noting that the bands were independent of the surface and share no relationship with the features of the surface.
This task is designed to observe the basic differences between the interaction modalities as force-based haptics provides a feel of an opposing force from the surface which may aid or hinder movement on the surface.

\textbf{\texttt{Visual Groove Brushing}} encompasses brushing curves on visually highlighted grooves on a surface (see \Cref{fig:groove}). The grooves coincide with the valleys formed when multiple protruded regions come together. We generated surfaces using method one described in \cref{sec:surf_models} and manually highlighted the grooves with bands to assist in visual querying and movement of the interaction modality on the surface. When using the haptic stylus, the grooved regions impart forces that guide the stylus to stay within the grooves. This task is therefore designed to observe if there is any advantage to using force-based haptics when the surface itself indirectly provides haptic assistance.   

\textbf{\texttt{Visual Annotation Brushing}} consists of brushing annotating curves on the surface (see \Cref{fig:annotation}). The visually highlighted annotation regions here represent feature formations on the surfaces. We acquired these surfaces through method three mentioned in \cref{sec:surf_models}. The surface models included a chair, a cup and a plane that provided a broad range of unique features for interaction. We manually highlighted the features on the surface with bands to guide the brushing of the annotation curves. Our goal here is to observe how interaction through force-based haptic facilitates or hinders movement on the surface when demarcating arbitrarily shaped intrinsic features of the surface.

For all \texttt{Brushing Curves} tasks, we captured the brushed curves by tinting the texels that corresponded to the positions of the laser pointer (haptic stylus or VR handheld) as it traversed the surface. Owing to the variable speed in the movement of the interaction modality, the pointer occasionally skipped texels during the brushing process. As a solution, we linked the previous and present pointer positions by coloring the intervening texels in a linear fashion. 


\section{User Study}

\subsection{Research Questions} \label{sec:res_qstns}
We formulate two research questions to evaluate the performance of using force-based haptics for immersive interaction visualization tasks on surfaces as compared to a non-haptic VR handheld controller. 
\begin{itemize}
    \item What is the comparative relationship between completion time and accuracy for \texttt{Point Localization} tasks performed on immersive surface visualizations using two different interaction modalities: a force-based haptic stylus (\texttt{HAPTIC}) and a handheld VR controller (\texttt{VR\_HAND})?
    \item What is the comparative relationship between completion time and accuracy for \texttt{Curve Brushing} tasks performed on immersive surface visualizations using the \texttt{HAPTIC} and \texttt{VR\_HAND} interaction modalities?
\end{itemize}

\subsection{Study Design}
The study followed a between-subject design, where participants were exposed to only one type of interaction modality to perform the interaction tasks enumerated in \cref{sec:int_des_rat}. 

\textbf{Tasks/Trials}:
The \texttt{Point Localization} and \texttt{Brushing Curves} tasks outlined in \Cref{sec:int_des_rat} are hereinafter referred to as \texttt{TaskProtrusion}, \texttt{TaskDepression}, \texttt{TaskRandVal}, and 
\texttt{TaskCurve}, \texttt{TaskGroove},  \texttt{TaskAnnotation} respectively. To account for variability in surface visualizations, each of the six tasks was further divided into three trials consisting of realizations with different levels of task complexity. 
Each participant performed a total of $18$ trials, $3$ trials per task. The order of the tasks and trials was counterbalanced using a balanced Latin square, whereas the assignment of the interaction modality was randomized.

\textbf{Datasets}: We generated the surface visualizations for each task with the methods outlined in \cref{sec:surf_models}. The surface models were generated through Python scripts and Blender.

\begin{figure}[tb]
  \centering 
  \includegraphics[width=\columnwidth]{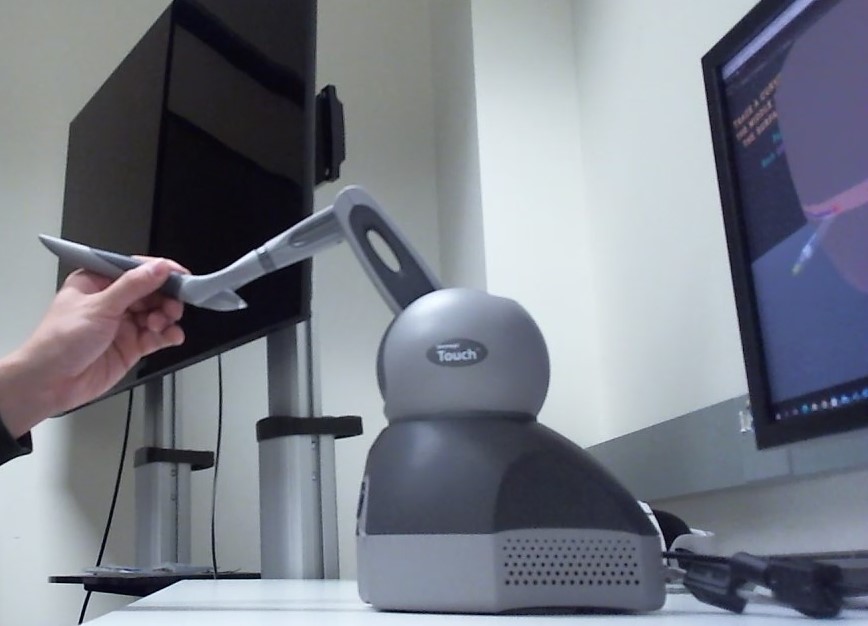}
  \caption{%
  	A left-handed participant using the force-based haptic device for interacting with surface visualizations.
  }
  \label{fig:participant_haptic}
\end{figure}

\textbf{Apparatus}: The modalities for interaction were implemented in Unity. The trials were conducted on a desktop computer (AMD Ryzen 7, 32 GB of RAM and GeForce RTX3060) with an Oculus Rift S headset which provides a comfortable VR experience. We utilized the GeoMagic\textsuperscript{\sffamily\textregistered} Touch\textsuperscript{\sffamily\texttrademark} Professional haptic stylus for a realistic and accurate experience for the force-based haptic sensation. The sessions were conducted using a seated VR setup with the haptic stylus placed on the dominant hand side of the participants. \Cref{fig:participant_haptic} shows a picture of a participant using \texttt{HAPTICS} for one of the trials.

\textbf{Participants}: We recruited a total of $40$ participants ($15$ females and $25$ males) by advertising the study to undergraduate and graduate students, faculty and postdoctoral fellows. Age ranged from $19$ to $47$ (mean=$26.5$, SD=$5.06$), 2 participants were left-handed. 
Prior experience using VR headsets in the context of 3D mapping applications such as Google Earth VR was used as a recruitment criterion.
Participants were also screened through a VRISE test~\cite{vrise_sharples} (score < 25) to avoid the effects of nausea and cybersickness in the study. Prior experience with haptic devices was not required. 

\subsection{Procedure}
Our study sessions comprised three phases: introduction and tutorial, the main trials, and finally, the post-study questionnaire. Most participants took $60$ to $90$ minutes to complete their session. In the introduction and tutorial phase, we explained the purpose of conducting the study. Then, the participants were guided through a training session where they gained familiarity with the type of interaction tasks and the modality (\texttt{HAPTIC} or \texttt{VR\_HAND}). Participants were encouraged to ask questions and take time to get comfortable with the interaction modality and the controls after which, the main trials started. 
For each trial, they followed instructions on the screen to perform the task outlined and submit their result when they felt that they had completed the localization or brushing task. The participants were instructed to complete the trials quickly and accurately.

 Before each trial, completion and submission instructions were shown on a panel including information regarding crucial control buttons specific to the trial. The participants could maneuver the visualization using a VR handheld controller in their non-dominant hand while performing point localization or curve brushing interactions using their dominant hand with the assigned modality.   
Once they felt confident about completing the requirements of a trial, they pressed the Joystick Button on their non-dominant hand controller to submit and move on.

Lastly, we administered a post-study questionnaire to the participants to obtain their feedback on the visualizations and the interaction modalities employed in the study, which included a series of multiple-choice questions and an opportunity to provide any suggestions or comments.

\subsection{Measures}
\label{sec:measures}
We measured two primary objective quantities for each trial. The first quantity is the \textbf{task completion time} and measures the time elapsed from the initiation of the trial until the participant clicks the submit button. The second quantity is the actual submitted \textbf{user response} and consists of the localized point value and the brushed curve textures for the \texttt{Point Localization} and \texttt{Brushing Curves} tasks respectively. During each trial, we also logged several additional quantities on a per-frame basis including the position of the controllers, the interaction modality employed, the navigation transformation events, the touch events on the surface from the \texttt{HAPTIC} modality, and the selection, draw, and erase events. We recorded the brushed curves from the participants by storing the corresponding brush texture from the surface mesh, along with logging scene information, such as positions and UV coordinates on the surface mesh for the brushed curves. We also video-taped the participants and screen-recorded their trials.


\section{Results} \label{sec:results}
We now report, interpret and discuss our findings on the performance of using \texttt{HAPTIC} and \texttt{VR\_HAND} for interacting with surface visualizations. We applied estimation techniques to the data collected from the 720 trials (\cref{sec:measures}). We report medians with confidence intervals (CI) following the recommended practises \cite{inference_by_eye_cumming, fair_stat_comm_hci_dragicevic} and avoid p-value statistics \cite{APA, the_new_stat_geoff}, which can still be obtained from the data following similar techniques. We refer the reader to the supplementary material for the compiled data. We used the median for statistical inference as it highlights the true center of the distribution which is not affected by outliers. We utilized pairwise differences between medians
and their 95\% CIs for our inferential analysis, indicating range of plausible values for the population median. We used empirical bootstrapping to construct the confidence intervals. 

We first present a holistic overview of the performance of each interaction modality before delving into a more in-depth analysis of some of the more intricate aspects.

\subsection{Overview}
\begin{figure}[tb]
  \centering 
  \includegraphics[width=\columnwidth, trim={3mm 6mm 3mm 6mm},clip]{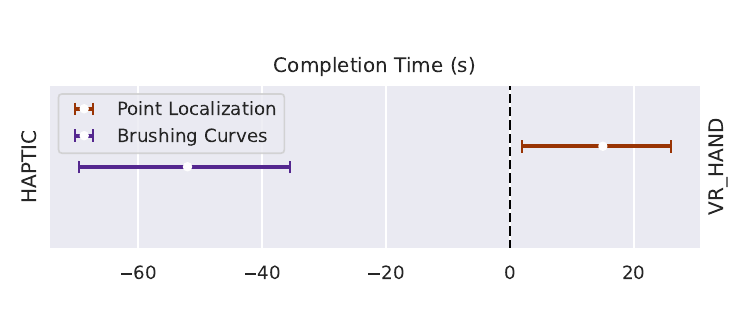}
  \caption{%
  	A holistic comparison between \textcolor{point_color}{\texttt{Point Localization}} and \textcolor{brush_color}{\texttt{Brushing Curves}} tasks with respect to total time taken (median) for completion. Significant differences are highlighted with a \textcolor{sig_color}{white} marker. 
  }
  \label{fig:overall_task_times}
\end{figure}

\begin{figure}[tb]
  \centering 
  \includegraphics[width=\columnwidth, trim={3mm 6mm 3mm 6mm},clip]{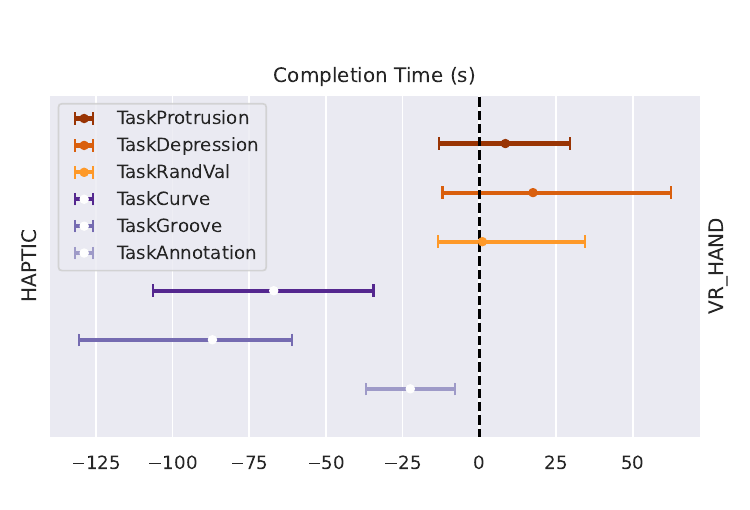}
  \caption{%
  	Task-wise comparison between \textcolor{point_color}{\texttt{Point Localization}} and \textcolor{brush_color}{\texttt{Brushing Curves}} tasks with respect to median completion time. 
  }
  \label{fig:per_task_times}
\end{figure}

Holistically, observing the pairwise difference between the two modalities, participants spent $15.0$s (median) (CI = $[26.0,\,2.0]$) longer when doing \texttt{Point Localization} tasks using \texttt{VR\_HAND} as compared to  \texttt{HAPTIC}, whereas participants spent $52.0$s $[-35.5,\,-69.5]$ more when using \texttt{HAPTIC} for \texttt{Brushing Curves} tasks compared to \texttt{VR\_HAND} (see \cref{fig:overall_task_times}). This pairwise comparison of \texttt{HAPTIC} and \texttt{VR\_HAND} shows strong evidence that the former is faster in \texttt{Point Localization} tasks and slower in \texttt{Brushing Curves} tasks compared to its counterpart. On a more granular level, under \texttt{Point Localization} tasks, \texttt{VR\_HAND} took $8.5$s $[29.5,\,-13.0]$ $17.5$s $[62.5,\,-12.0]$, and $1.0$s $[34.5,\,-13.5]$ more than \texttt{HAPTIC} for \texttt{TaskProtrusion}, \texttt{TaskDepression} and \texttt{TaskRandVal} respectively, but the evidence suggests that the task-wise differences were not significant. However, for \texttt{Brushing Curves} tasks, \texttt{HAPTIC} took $67.0$s $[34.5,\,-106.5]$, $87.0$s $[-61.0,\,-130.5$], and $22.5$s $[-8.0,\,-37.0]$ longer than \texttt{VR\_HAND} for \texttt{TaskCurve}, \texttt{TaskGroove}, and \texttt{TaskAnnotation} respectively. \cref{fig:per_task_times} shows  evidence that the differences between the two interaction modalities were significant. 

\begin{figure}[t]
  \centering 
  \includegraphics[width=\columnwidth, trim={3mm 6mm 3mm 6mm},clip]{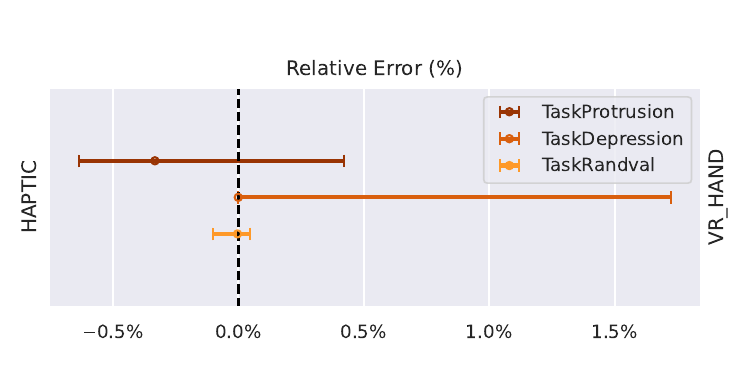}
  \caption{%
  Comparison of relative error for \texttt{\pointc{Point Localization}} tasks.
  }
  \label{fig:per_task_relative_error}
\end{figure}

\begin{figure}[t!]
  \centering 
  \includegraphics[width=\columnwidth, trim={3mm 6mm 3mm 6mm},clip]{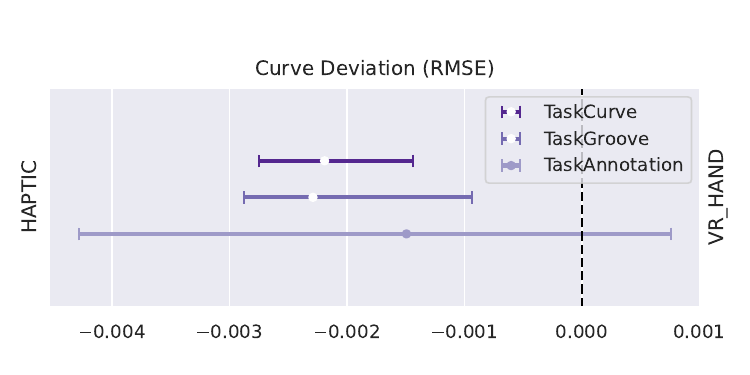}
  \caption{%
  Comparison of Curve Deviation for  \texttt{\brushc{Brushing Curves}} tasks. 
  }
  \label{fig:per_task_curve_divergence}
\end{figure}

We now move on to analyzing the error in \texttt{Point Localization} and \texttt{Brushing Curves} tasks using both modalities at a per task level. For \texttt{Point Localization}, we computed the relative error as compared to the ground-truth localization point. We observed that there was no significant difference in relative error when localizing points using either modality; however, for \texttt{TaskDepression} $0\%$ $[0\%,\,1.72\%]$, a skewness towards \texttt{VR\_HAND} indicates lower relative error for \texttt{HAPTIC} but more data is needed to get a conclusive outcome (see \cref{fig:per_task_relative_error}).\\
For the \texttt{Brushing Curves} tasks, we quantified the deviation of the brushed curve from the medial axis of the visualized band that the participants were asked to follow. We first computed the Euclidean distance transform (EDT) on the brushed texture image and compared it to the EDT of the medial axis of the band texture shown on the visualization. Finally, the root mean square error (RMSE) was computed on the two images. The Curve Deviation is defined as
\begin{equation}
\Bigl[
\frac{1}{\lvert \mathcal{B} \rvert}
\sum_{(i,j) \in \mathcal{B} } 
\bigl(
D_{\text{G}}[i,j] - D_{\text{X}}[i,j]
\bigr)^2
\Bigr]^{\tfrac{1}{2}},
\label{eqn:curve_deviation}
\end{equation}
where $D_\text{G}$ is the EDT of the reference medial axis texture, $D_\text{X}$ is the EDT of the brushed texture ($\text{X}$ is either \texttt{HAPTIC} or \texttt{VR\_HAND}), and $\mathcal{B}$ is the set of texels that fall within the band shown to the participants. 
We observed that the curve deviation when using \texttt{HAPTIC} was higher than \texttt{VR\_HAND} in \texttt{TaskCurve} ($-0.0022$ $[-0.0014,\,-0.0027]$) and \texttt{TaskGroove} ($-0.0023$ $[-0.0009,\,-0.0029]$), but no significant difference was found for \texttt{TaskAnnotation} (see \cref{fig:per_task_curve_divergence}). 

\subsection{Further Analysis}

\subsubsection{Point Localization}
\begin{figure}[t]
  \centering 
  \includegraphics[width=\columnwidth, trim={3mm 6mm 3mm 6mm},clip]{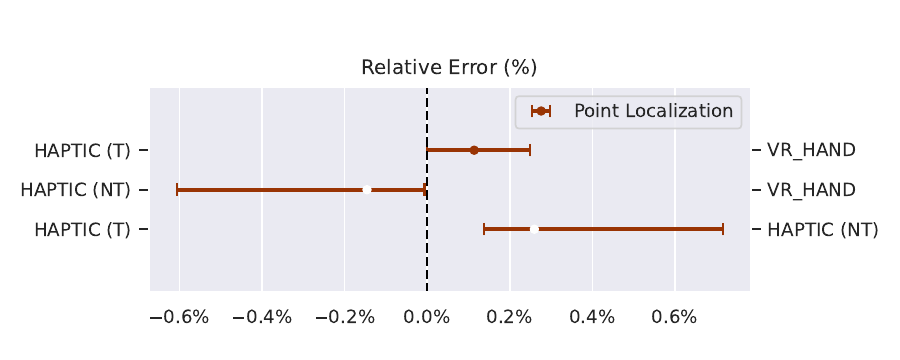}
  \caption{The difference in relative error for \pointc{Point Localization} tasks when the user is touching the surface: \texttt{HAPTIC (T)}, not touching the surface: \texttt{HAPTIC (NT)} and using visual-only mode: \texttt{VR\_HAND}.}
  \label{fig:touch_no_touch_vr_relative}
\end{figure}

\begin{figure}[t]
  \centering 
  \includegraphics[width=\columnwidth, trim={3mm 6mm 3mm 6mm},clip]{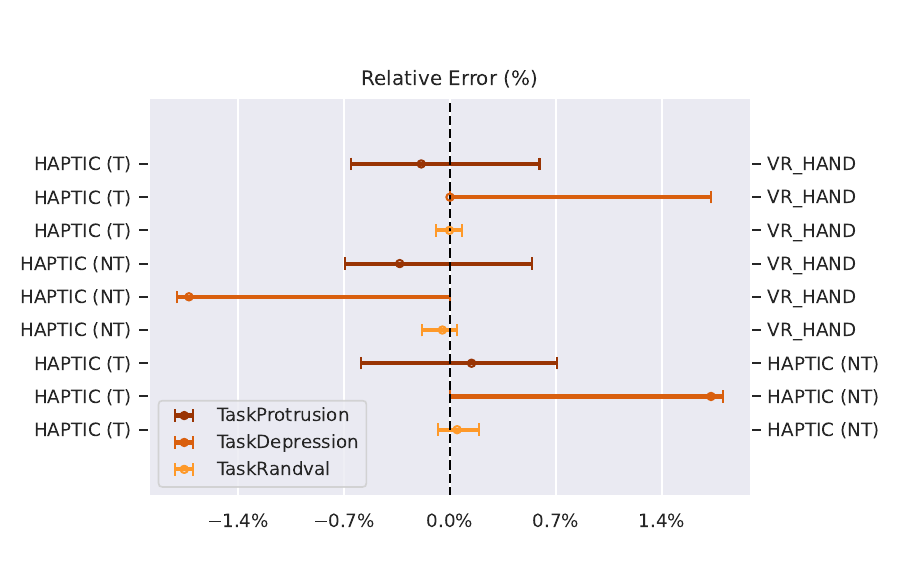}
  \caption{Per-task difference of relative error for \pointc{Point Localization}. 
  }
  \label{fig:per_task_touch_no_touch_vr_relative}
\end{figure}

At a high level, we observed no significant difference between the two modalities in terms of accuracy when it comes to \texttt{Point Localization}. This is counter-intuitive as the \texttt{HAPTIC} modality allows a user to physically feel the surface thereby facilitating interpretation and cognition. We observed that some participants were relying more on visual feedback and not touching the surface when submitting their responses. We, therefore, inspected the relative error made when the user was indeed touching the surface before submitting (\texttt{HAPTIC (T)}), as compared to the cases where they were not touching (\texttt{HAPTIC (NT)}) or using \texttt{VR\_HAND}, and observed that \texttt{HAPTIC (NT)} was significantly more error-prone ($0.25\%$ $[0.71\%,\,0.13\%]$) as compared to \texttt{HAPTIC (T)}. \texttt{HAPTIC (NT)} was also more error-prone compared to \texttt{VR\_HAND} ($-0.145\%$  $[-0.006\%,\,-0.606\%]$. However, when we compared \texttt{HAPTIC (T)} against \texttt{VR\_HAND}, the participants comparatively had a higher relative error with the latter ($0.11\%$ $[0.24\%,\,0.00\%]$) (see \cref{fig:touch_no_touch_vr_relative}). These results highlight that when a participant physically made contact with the surface visualization for localizing a point, they were less prone to error as compared to relying on the visual-only stimuli. At a per-task level, we observed that although there were no significant differences between the interaction modalities, using \texttt{VR\_HAND} for \texttt{TaskDepression} was still comparatively more error-prone as compared to the \texttt{HAPTIC (T)} which indicates that the physical touch sensation assisted in localizing points in regions where it was difficult to perceive depth visually (see \cref{fig:per_task_touch_no_touch_vr_relative}). 


\begin{figure}[t!]
  \centering 
  \includegraphics[width=\columnwidth, trim={3mm 6mm 3mm 6mm},clip]{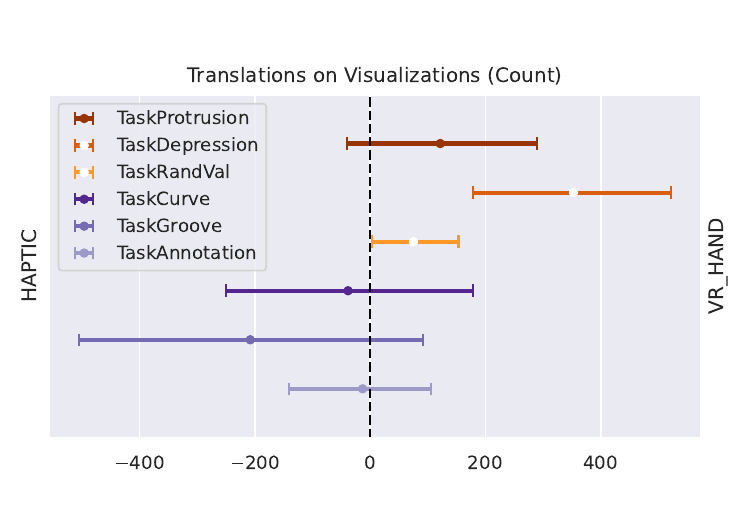}
  \caption{A difference between the number of translations performed for every \pointc{Point Localization} and \brushc{Brushing Curves} tasks. 
  }
  \label{fig:per_task_translations}
\end{figure}

\begin{figure}[t]
  \centering 
  \includegraphics[width=\columnwidth, trim={3mm 6mm 3mm 6mm},clip]{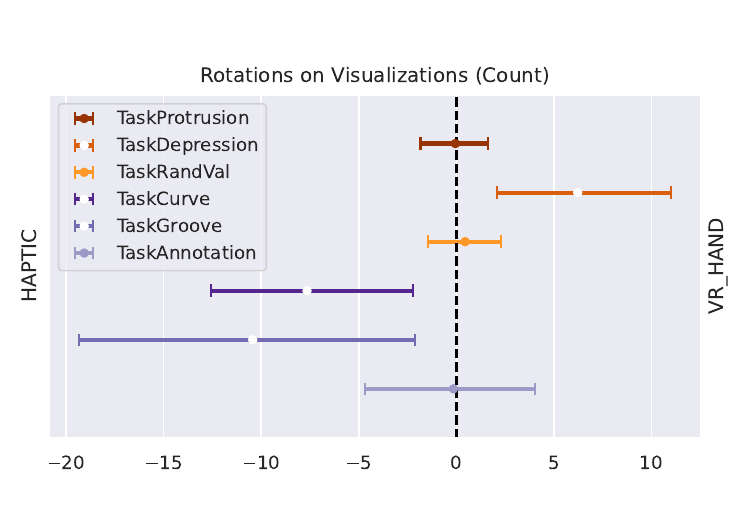}
  \caption{The difference in number of rotations performed for every \pointc{Point Localization} and \brushc{Brushing Curves} tasks. 
  }
  \label{fig:per_task_rotations}
\end{figure}

We further examined the number of navigation transformations that the participants performed during the tasks and observed that there were significantly more translations ($352.8$ $[522.1,\,178.1]$) and rotations ($6.2$ $[11.0,\,2.1]$) during \texttt{TaskDepression} when using \texttt{VR\_HAND}; we noticed a similar trend in \texttt{TaskRandVal} but only for the number of translations ($75.2$ $[153.2,\,2.4]$) (see \cref{fig:per_task_translations} and 
\cref{fig:per_task_rotations}). The increased number of transformations for the \texttt{TaskDepression} asserts our earlier findings that in scenarios that require depth perception with visual-only simuli, more transformations of the visualizations were required to compare and contrast different regions, that eventually reflects on the time taken and the accuracy of localization. However, 
we observed no significant differences between the modalities for \texttt{TaskRandVal} and \texttt{TaskProtrusion}. For \texttt{TaskRandVal}, we attribute this to the quality of the visual stimulus provided; the surface has visually distinct regions that likely guided the participants toward the correct value thereby diminishing the need for haptic assistance. 
Similarly, in \texttt{TaskProtrusion}, the protrusions on the surface are visually salient; the participants likely relied on strong visual cues to gauge the protrusion level and localize the highest point on the surface.


\subsubsection{Brushing Curves}
\begin{figure*}[h]
  \centering
  \begin{subfigure}[b]{0.23\linewidth}
  	\centering
  	\includegraphics[width=\textwidth]{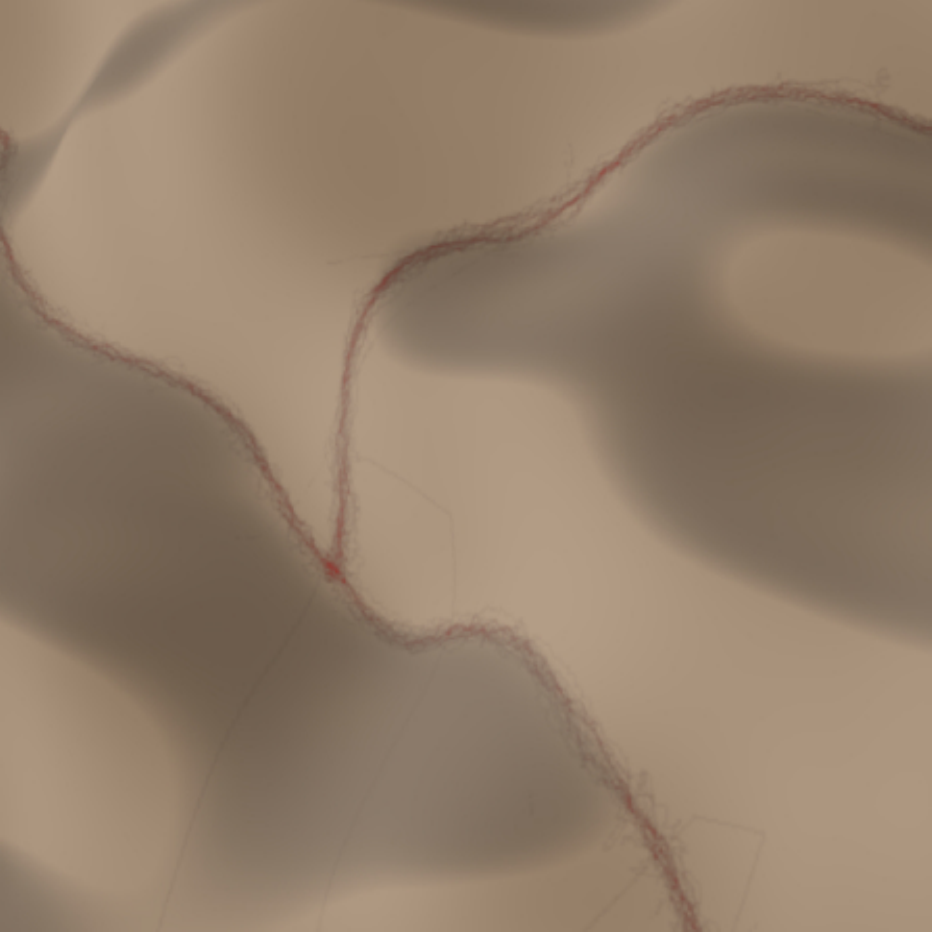}
  	\caption{\texttt{VR\_HAND} Draw Events (Groove)}
  	\label{fig:vde1}
  \end{subfigure}%
  \hfill%
  \begin{subfigure}[b]{0.23\linewidth}
  	\centering
  	\includegraphics[width=\textwidth]{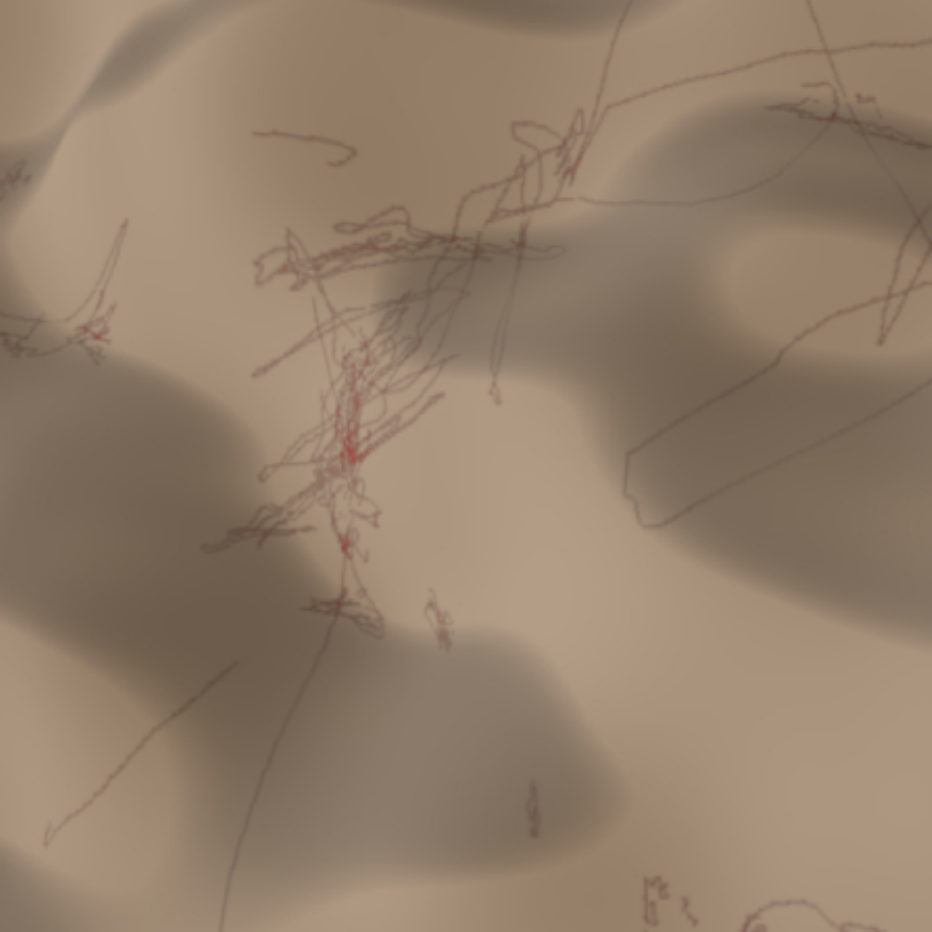}
  	\caption{\texttt{VR\_HAND} Erase Events (Groove)}
  	\label{fig:vee1}
  \end{subfigure}%
  \hfill
  \begin{subfigure}[b]{0.23\linewidth}
  	\centering
  	\includegraphics[width=\textwidth]{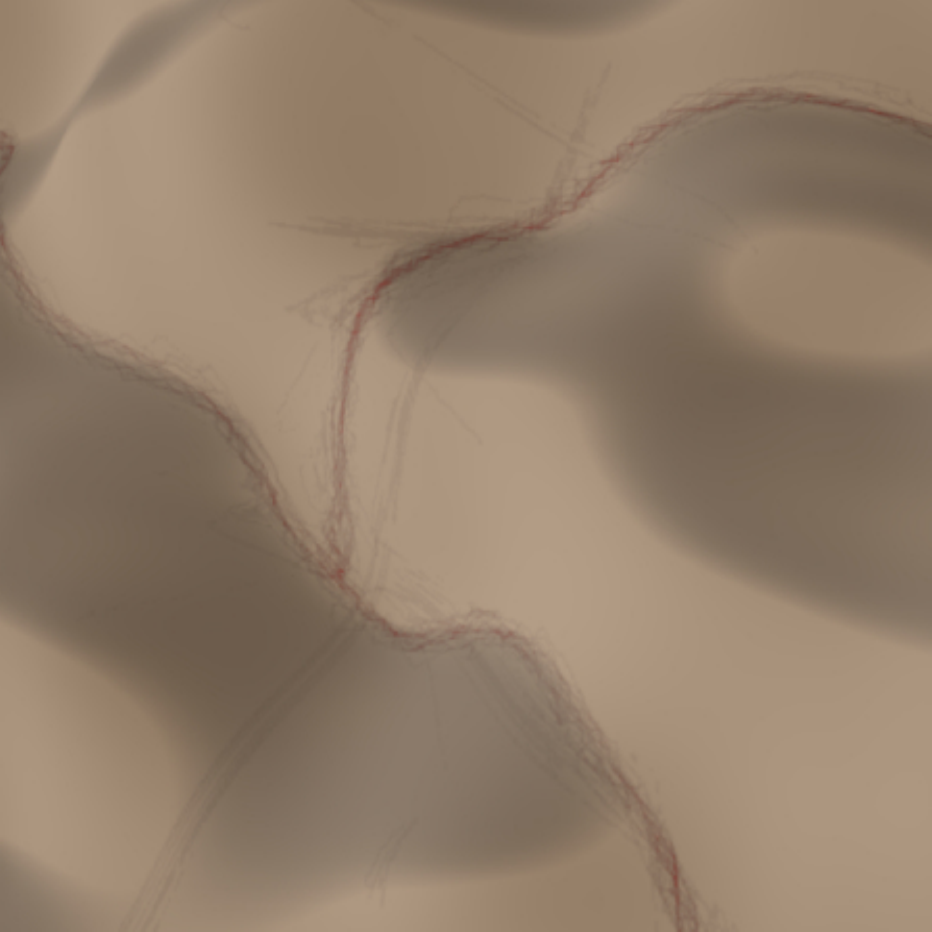}
  	\caption{\texttt{HAPTIC} Draw Events (Groove)}
  	\label{fig:hde1}
  \end{subfigure}%
  \hfill
  \begin{subfigure}[b]{0.23\linewidth}
  	\centering
  	\includegraphics[width=\textwidth]{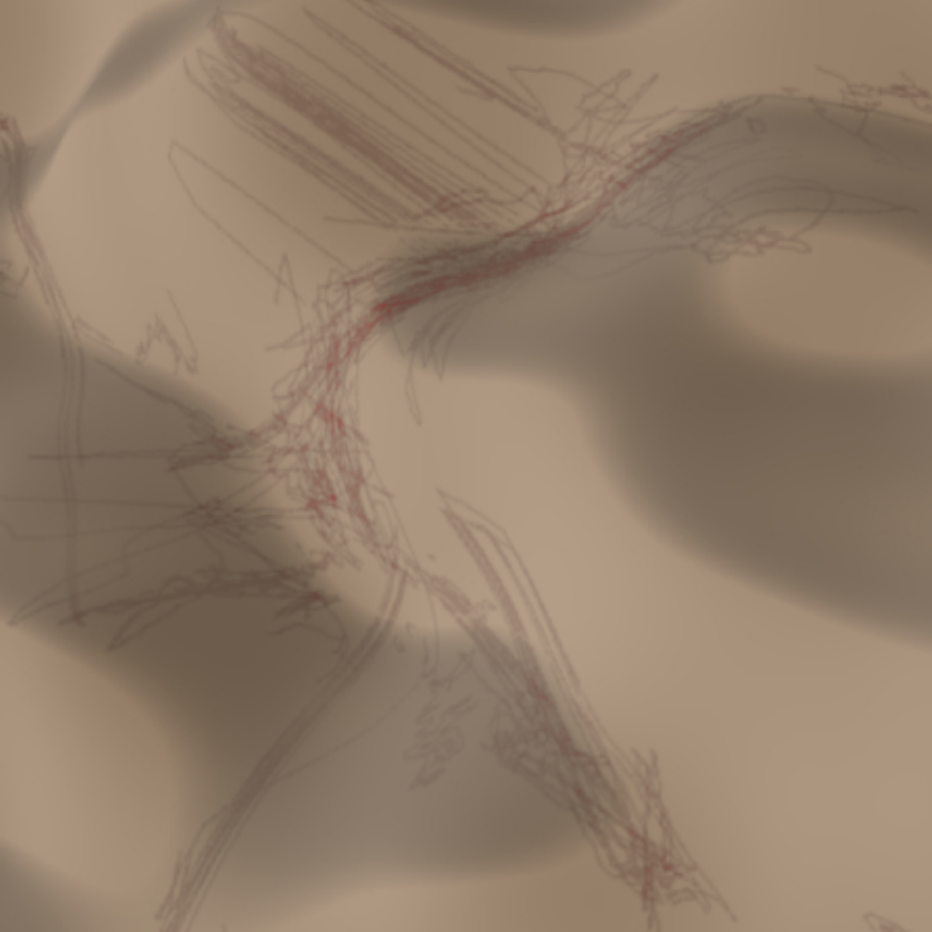}
  	\caption{\texttt{HAPTIC} Erase Events (Groove)}
  	\label{fig:hee1}
  \end{subfigure}%
  \subfigsCaption{Heatmap of curves brushed by the participants and the corresponding corrections made on the surface visualizations when using \texttt{HAPTIC} and \texttt{VR\_HAND} respectively. }
  \label{fig:curves_heatmap}
\end{figure*}

In our analysis of \texttt{Brushing Curves} tasks, we observed that the participants took longer using \texttt{HAPTIC} as compared to \texttt{VR\_HAND}, and also brushed curves that deviated more from the medial axis.
During the trials, we instructed the participants to stay within the visualized band on the surface and make corrections when the curve leaves the bounds of the band. We further instructed them to brush the curves as close to the center of the band as possible. Upon inspection of the curves drawn by the participants, we found that when the participants were brushing on a region with the stylus slanted away from the normal, 
the normal force pushed the \texttt{HAPTIC} stylus away from the medial axis; this required the participants to apply a counterbalancing force to trace within the band (see \cref{fig:curves_heatmap}). In these regions, we also noticed that the \texttt{HAPTIC} participants made more corrections. 
Moreover, rapidly changing topography of the surface visualization impacts brushing tasks for both \texttt{HAPTIC} and \texttt{VR\_HAND}; a slight perturbation of the controller leads to a rapid change in the position of the laser pointer resulting in errors
(see \cref{fig:vee1} and \cref{fig:hee1}). 


\begin{figure}[tbh]
  \centering 
  \includegraphics[width=\columnwidth, trim={3mm 6mm 3mm 6mm},clip]{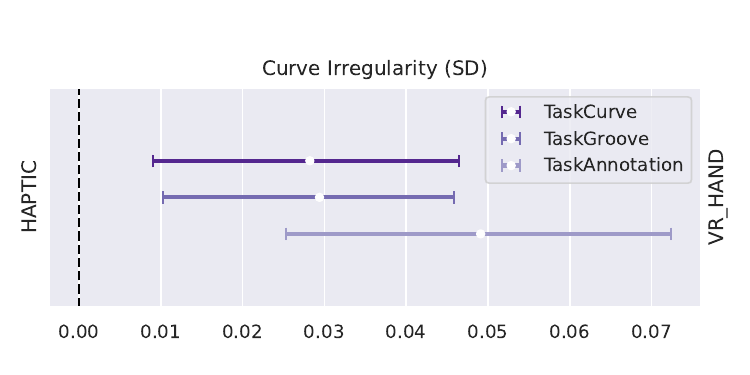}
  \caption{A task-wise comparison of Curve Irregularity (SD) between \texttt{HAPTIC} and \texttt{VR\_HAND} for \brushc{Brushing Curves} tasks. Significant differences are highlighted with \sigc{white} markers. \hamza{correct the STD}}
  \label{fig:per_task_curve_irregularity}
\end{figure}

\begin{figure*}[t!]
  \centering 
  \includegraphics[width=\linewidth, trim={0 14.6cm 0 0}, clip]{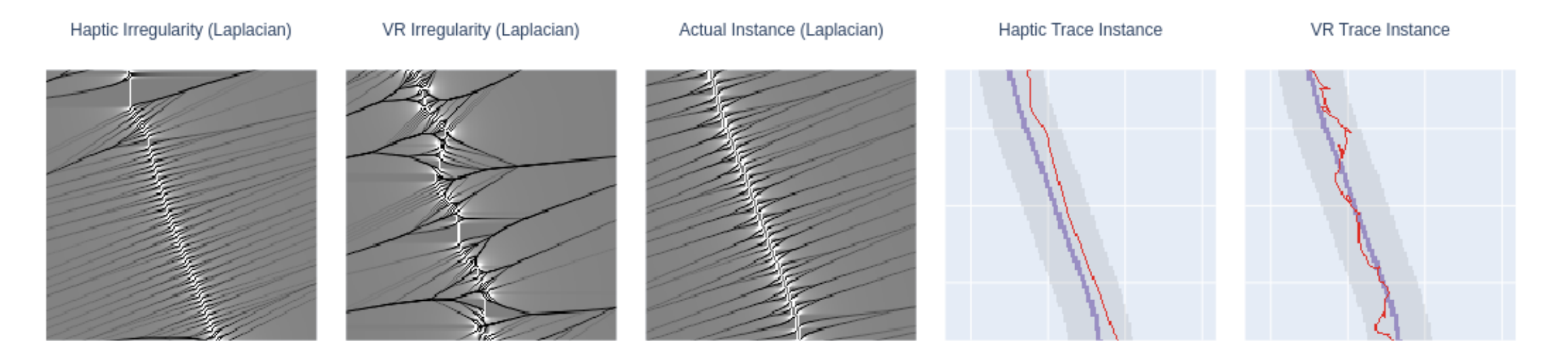}
  \caption{A comparison of curve smoothness and deviation from the medial axis between curves brushed using \texttt{HAPTIC} and \texttt{VR\_HAND}.}
  \label{fig:curve_smoothness}
\end{figure*}
We also noticed a remarkable difference in the regularity of the curves drawn using the two interaction modalities. To further quantify this, we computed the EDT on the brushed texture image 
(\cref{eqn:curve_deviation}), followed by a Laplacian of the EDT. Finally, we computed the standard deviation (SD) of the Laplacian to characterize the irregularity of the curved. Specifically, the Curve Irregularity (SD) is defined as
\begin{equation}
\mathrm{{SD}} \bigl[
\left\{\nabla^2{D_X}[i,j] : (i,j) \in \mathcal{B} 
\right\}
\bigr]
,
\label{eqn:curve_irregularity}
\end{equation}
and measures jitter in the curves which is attributable to hand stability when brushing on the surfaces. When using \texttt{VR\_HAND}, the curves brushed by the participants were significantly more irregular than the curves brushed using \texttt{HAPTIC} in all three tasks (see \Cref{fig:per_task_curve_irregularity}). \Cref{fig:curve_smoothness} visually depicts the irregularity of the curves drawn by \texttt{HAPTIC} and \texttt{VR\_HAND} respectively. Here, we observe that the \texttt{HAPTIC} curves were smoother but deviated from the medial axis, whereas, though the curves brushed by \texttt{VR\_HAND} were closer to the medial axis, they exhibited higher irregularity.


Additionally, we analyzed the transformations performed on the \texttt{Brushing Curves} tasks. We observed that there was no significant difference in the number of translation events (see \cref{fig:per_task_translations}), however, we found that the participants were performing significantly more rotations during \texttt{TaskCurve} and \texttt{TaskGroove} when using \texttt{HAPTIC} (\cref{fig:per_task_rotations}). This shows that the participants using \texttt{HAPTIC} maneuvered the visualization more to trace the occluded regions of the visualizations, given that they needed to be in close proximity to the surface. 

\begin{figure}[tbh]
  \centering 
  \includegraphics[width=\linewidth, trim={0.3cm 0cm 5.5cm 0cm}, clip]{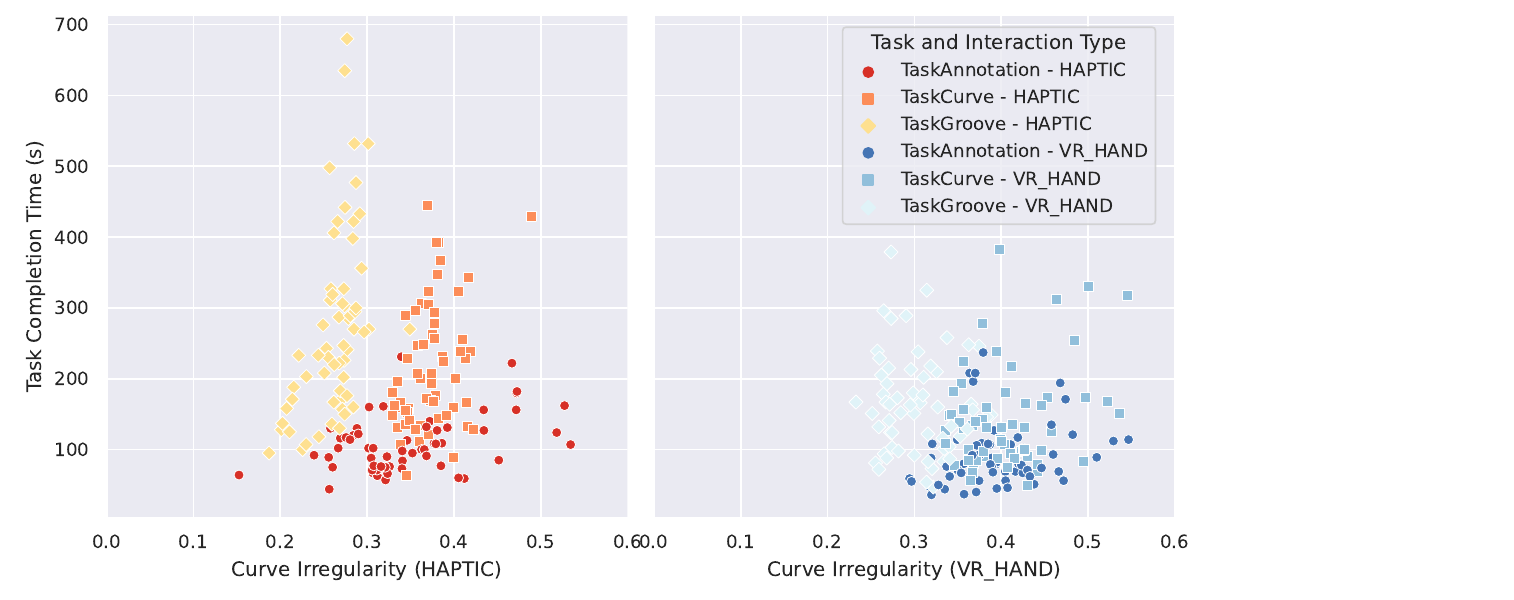}
  \caption{%
  Per-task correlation between Curve Irregularity and task completion time for \pointc{\texttt{HAPTIC}} and \brushc{\texttt{VR\_HAND}} on \texttt{Brushing Curve} tasks.
  }
  \label{fig:brushing_correlations}
\end{figure}

Finally, \cref{fig:brushing_correlations} shows the correlation between curve irregularity and completion time for both the interaction modalities for all the \texttt{Brushing Curves} tasks. We observed that for \texttt{TaskGroove}, the participants who took less time brushing curves using \texttt{HAPTIC} generated smoother curves and the curves started to become irregular as the time increased. On closer inspection, we also observed that the \texttt{HAPTIC} curves brushed for \texttt{TaskGroove} showed the least irregularity and we attribute this to the indirect assistance provided by the grooves. 



\section{Participants' Feedback}\label{sec:feedback}
\begin{figure}[t]
  \centering 
  \includegraphics[width=\columnwidth, trim={0cm -1cm 0cm 0cm}, clip]{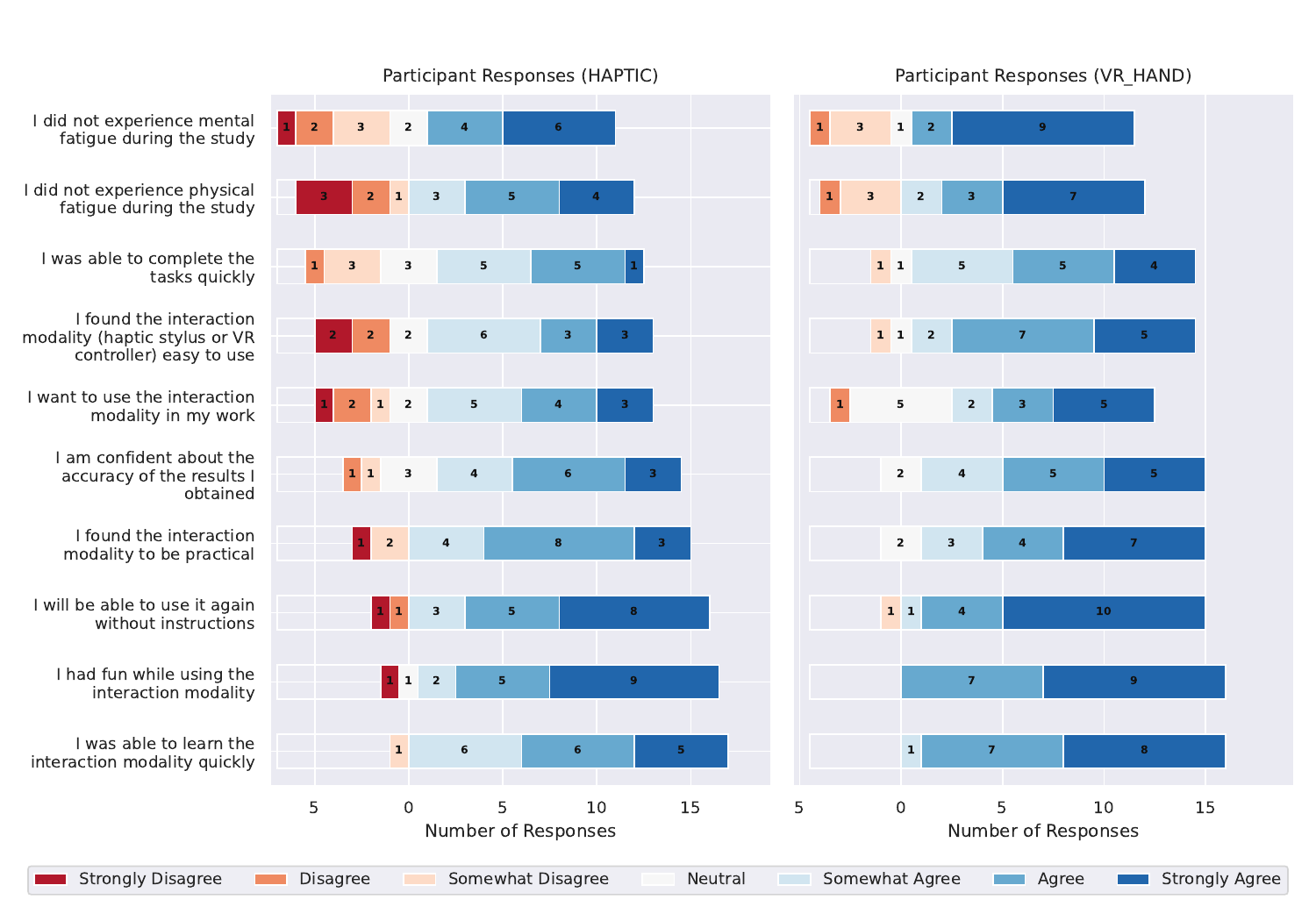}
  \caption{A Likert scale summarizing the participant responses from the post-study questionnaire on each of the interaction modalities.}
  \label{fig:participants_response}
\end{figure}

In the final phase of the study, we asked the participants to fill out a post-study questionnaire to gather feedback on their overall experience and suggestions for improvement. 
\Cref{fig:participants_response} shows the participant responses for both the interaction modalities. Overall, participants using \texttt{HAPTIC} reported higher levels of mental and physical fatigue. We attribute this to the participants' lack of experience with a force-based haptic device. 
Moreover, \texttt{HAPTIC} participants were comparatively less confident about the accuracy of their results. We speculate that this is attributable to a combination of factors that have to do with the participants' lack of experience and the errors induced because of the forces applied from the surface while brushing (see \cref{sec:results}). However, participants in both cases learned to use the interaction modality quickly and had a fun experience, and were confident that they would be able to use the modality again without any instructions. 

Additionally, participants shared comments on the interaction modalities and their experience while using it for interactions. While using \texttt{VR\_HAND} for interaction, participants
reported experiencing physical fatigue, e.g. \textit{"(...) The interaction modalities could weigh less to minimize physical fatigue."} (V1). Similarly, they reported their comfort level, e.g. \textit{"(...) it can be physically uncomfortable to wear/hold the interaction modalities for a long time"} (V2). Moreover, they reported on the learning curve of using the controls and interaction modality, for instance \textit{"(...) consider a learning time to get familiarize with the device and buttons"} (V3). A few of the participants had reservations on the accuracy, where one of them described it as \textit{"(...) it was difficult to select precise locations on the mesh and was quite difficult to select precise points and using the VR controller for the outlining tasks was difficult and I had to be very careful to `stay within the lines'"} (V4), whereas another participant mentioned it as \textit{"Small vibrations of the hand affected the result greatly"} (V5). Another participant reported on the accuracy of brushing curves using \texttt{VR\_HAND} as \textit{"The only problem that I experienced was the difficulty to draw lines smoothly because the hands are in the air without any support"} (V6).

Similarly, when using \texttt{HAPTIC} for interactions, participants reported experiencing physical fatigue and suggested measures to counteract it, such as \textit{"I felt like if I had a resting pillow under my right arm, it could be more easy for me to draw lines accurately."} (H1). Moreover, they reasoned that the effect on their performance resulting from their lack of experience using the force-based haptic device, for instance \textit{"I felt it was hard to get used to it at first. Practicing more might help."} (H2). Moreover, they reported the excessive need of navigation required for interacting with the visualization while brushing curves as \textit{"User have to navigate and adjust the terrain multiple times to complete the line tracing task"} (H3). Furthermore, they reported the impact of surface normal forces while brushing curves as \textit{"As it draws point to point, I several times drew a line unwanted on edge of surfaces"} (H4). Some of the participants commented on the ergonomics of the physical haptic stylus by suggesting a more comfortable holding position, where one quoted "(...) I would feel it more comfortable if I had not held the middle. (...)" (H5). Moreover, they also reported the performance impact from physically touching the surface when brushing, for instance \textit{"(...) When the pen was touching the surface and it was engaged, following the counters along the valleys was easier (...)"} (H6). Furthermore, they highlighted the negative effect of surface normal force on brushing curves in \texttt{TaskCurve} as \textit{"(...) I preferred to follow the contours without touching the surface in general (...)"} (H6).

\section{Takeaways}
In this section, we summarize our findings in the form of takeaways to provide design guidelines for future researchers using force-based haptics for surface visualization interactions. 

\textbf{Takeaway\#1:} \textit{Force-based haptics is relatively faster and less error-prone in localizing points on surface visualizations where perception of depth is involved}. Our analysis of the data and feedback from participants (V3) revealed a correlation that participants relying on physical feel from the surface for depth perception when localizing a point performed better than participants relying on visual-only stimuli. 

\textbf{Takeaway\#2:} \textit{Localizing points using visual cues only requires extensive maneuverability of the surface visualizations}. Our analysis highlights that participants who only  compared different surface regions visually performed more translations and rotations as compared to participants who additionally used the haptic stylus.

\textbf{\texttt{Takeaway\#3:}} \textit{Brushing curves using haptic force feedback requires extensive maneuverability of the surface visualization.} Our results and participants' feedback (H3) underlined that participants who brushed curves on the surfaces using the haptic stylus performed extensive rotations and movements to visually observe the regions due to the proximity requirement of the stylus.

\textbf{\texttt{Takeaway\#4:}} \textit{Accuracy of brushed curves using haptic force feedback is affected by the angle of the force applied in relation to the normal.} Our analysis and participants' feedback (H4 and H6) highlighted that participants applying a force on the stylus that deviates from the surface normal pushed the curves away from the intended brushing direction.

\textbf{\texttt{Takeaway\#5:}} \textit{Brushing on surface visualizations using haptic force feedback produces smoother curves.} Our analysis and feedback from participants (H6) showcased that the stability of the device and the assistance from the surface lead to smoother hand movements.

\section{Conclusion, Limitations and Future Work}
In this paper, we evaluated force-based haptic immersive interactions for surface visualizations against a visual-only mode of immersive interactions using VR handheld controllers. Our aim was to gauge the performance of the two modalities for surfaces visualization tasks that are obfuscated by occlusions and rely heavily on the quality of rendering for interpretation. We devised a user study involving tasks that encompass common interactions performed on surface visualizations: point localization and brushing curves. Based on the tasks, we formulated two research questions to evaluate the performance of haptic force feedback against the visual-only mode. Our results indicate that assistance from a haptic force feedback device allows faster localization of points that are occluded with respect to their surroundings (\texttt{TaskDepression}). The localization also requires less navigation and provides better accuracy. Furthermore, the force assistance from the surface results in smoother curves but the curves exhibit more deviation from the intended target. We also summarized feedback from participants and highlighted main takeaways from our work for future researchers. 

We present and acknowledge some limitations in our work and discuss avenues for future research. Although our quantitative study included participants who had considerable experience using VR handheld controllers, only two participants had experience with using a force-based haptic stylus. Our aim was to target a more generalized analysis of the efficacy of using the interaction modalities by recruiting participants from different backgrounds. Hence, we believe that participants not having experience with a haptic force feedback device can bias the results towards interactions performed using VR handheld controllers. Moreover, we also acknowledge that there is a high learning curve associated with using the interaction modalities and hence in future, long term effects on performance of participants using the interaction modalities can be studied in a longitudinal study. We further acknowledge that even though we observed significant differences between the two modalities for some tasks, some tasks showed almost significant results and can benefit from additional analysis with a higher number of participants.  


While our study involved a force-based haptic device with a stylus design which can be generalized to other devices that provide a similar force-based haptic sensation, we acknowledge that our study did not evaluate other forms of force-based haptic devices such as haptic gloves and other state-of-the-art devices. The evaluation of these devices is a subject of future research.
We used a combination of explicit surface models that were procedurally generated; we believe that these surface models are representative of the vareity of surfaces that users are likely to encounter for visualization purposes. However, future work could consider using other forms of surface models such as implicit surface visualizations to evaluate the interaction modalities. Additionally, we opted to limit our study to point localization and brushing curves tasks. Future work can explore a broader spectrum of haptic interaction tasks such as multivariate data encoding on surfaces and interactions with deformable surfaces as well as volumes.

Furthermore, our study focused only on the force based stimulus from the haptic device which provides a physical feel of the surface's shape and depth. Future work could consider exploring additional forms of force-based data haptifications such as force vibrations and surface friction for relaying multiple data encodings in the form of multivariate surface visualizations. We also acknowledge that our study did not consider the quality of the triangulation of the surface meshes; bad triangulations may impact the performance of interaction modalities. Finally, we also acknowledge that we only evaluated passive force based haptic assistance; future work can evaluate the performance of active force-based haptic assistance for surface visualization interactions.  




\bibliographystyle{abbrv-doi-hyperref}

\bibliography{template}

\end{document}